\documentclass[12pt]{article}
\usepackage{amsmath,amssymb}
\usepackage{jheppub}
\newdimen\tableauside\tableauside=1.0ex
\newdimen\tableaurule\tableaurule=0.4pt
\newdimen\tableaustep
\def\phantomhrule#1{\hbox{\vbox to0pt{\hrule height\tableaurule width#1\vss}}}
\def\phantomvrule#1{\vbox{\hbox to0pt{\vrule width\tableaurule height#1\hss}}}
\def\sqr{\vbox{%
		\phantomhrule\tableaustep
		\hbox{\phantomvrule\tableaustep\kern\tableaustep\phantomvrule\tableaustep}%
		\hbox{\vbox{\phantomhrule\tableauside}\kern-\tableaurule}}}
\def\squares#1{\hbox{\count0=#1\noindent\loop\sqr
		\advance\count0 by-1 \ifnum\count0>0\repeat}}
\def\tableau#1{\vcenter{\offinterlineskip
		\tableaustep=\tableauside\advance\tableaustep by-\tableaurule
		\kern\normallineskip\hbox
		{\kern\normallineskip\vbox
			{\gettableau#1 0 }%
			\kern\normallineskip\kern\tableaurule}%
		\kern\normallineskip\kern\tableaurule}}
\def\gettableau#1 {\ifnum#1=0\let\next=\null\else
	\squares{#1}\let\next=\gettableau\fi\next}

\begin{document}

%\today
\preprint{Imperial-TP-2023-CH-03}

\title{Magnetic  Charges for the Graviton}
\author{C.M. Hull}

\affiliation{The Blackett Laboratory, Imperial College London, Prince Consort Road, London SW7 2AZ, United Kingdom}

\emailAdd{c.hull@imperial.ac.uk}

\abstract{ Symmetries and conserved charges are investigated for linearised gravity and its dual formulation in terms of the dual graviton field. 
Conserved  charges are constructed for the
dual graviton theory that are associated with invariances of the dual graviton theory.
These invariances arise for gauge parameters that are certain generalised Killing tensors.
These electric-type charges for the dual graviton  are then shown to give magnetic-type charges for the graviton.
One of the magnetic-type gravitational charges obtained in this way is the linearisation of the gravitational charge  in $d\ge 5$ dimensions that arises  as a central charge in the supersymmetry algebra and is carried by Kaluza-Klein monopoles.  Solutions of linearised gravity carrying the magnetic gravitational charges are discussed.
The application of the approach used here to other gauge theories is discussed.}

%\keywords{}
                              
\maketitle

\section{Introduction and Overview}

One way of understanding magnetic charge in electromagnetism or antisymmetric tensor gauge theories is as the electric charge of the dual theory.
In this paper  magnetic-type charges will be introduced for linearised gravity as the electric-type charges of the dual formulation in terms of the dual graviton \cite{Hull:2000zn}.
The corresponding magnetic charges for  non-linear gravity 
  will   be discussed elsewhere. In this introduction, an overview of the construction and the results will be given, while the remainder of the paper will fill in the details.

Magnetic charge for gravity plays an important role in M-theory. Kaluza-Klein monopoles are an essential part of the spectrum of BPS states in M-theory \cite{Hull:1994ys} and the charge that they carry can be viewed as a magnetic gravitational charge.  On toroidal compactification, the gravitational field gives vector gauge fields (graviphotons)
and the magnetic monopoles for these graviphotons arise from Kaluza-Klein monopoles in the higher dimensional theory. Then
 the magnetic charge for the graviphotons should arise from a charge carried by the Kaluza-Klein monopoles. The BPS charge carried by Kaluza-Klein monopoles is a $d-5$ form charge that was introduced in \cite{Hull:1997kt}, 
where it was
 shown to agree with the central term in the superalgebra for Kaluza-Klein monopole solutions. (The explicit form of this charge is given in equation (\ref{kcharge}).)
  This charge can be viewed as a gravitational   magnetic charge that arises in dimensions $d\ge 5$, while the standard ADM momentum and angular momentum \cite{Abbott:1981ff} can be regarded as gravitational   electric charges. It was seen in \cite{Hull:2000zn,Hull:2001iu} that the Kaluza-Klein monopole charge can be thought of as a charge for the dual graviton.

 The aim of the present work is to further investigate such charges in gravity. In this paper it will be seen that the magnetic-type charges for linearised gravity that arise as electric-type charges for the dual graviton include the Kaluza-Klein monopole charge  that  appears  in the superalgebra but also include further charges. 
The case of 4 dimensions is rather different from that of $d>4$ -- for example, Kaluza Klein monopoles and their associated magnetic charges only arise for $d>4$. 
In the full non-linear theory, turning on magnetic charges such as the Kaluza Klein monopole charge can change the topology of the spacetime, 
whereas in the linear theory of fluctuations about a Minkowski metric such complications can be avoided.

Consider first the case of
a $(p + 1)$-form field strength $F$ in $d$ dimensional Minkowski space  
satisfying the  equations
\begin{equation}
{dF} = \ast \tilde{j}  \qquad d \ast F = \ast j
\label{Ffida}
\end{equation}
where
$j$ is  a   $p$-form electric
current  and $\tilde{j}$ is a $( d - p - 2)$-form magnetic current; these currents are conserved, $d \ast  {j}=0$, $ d \ast \tilde{j}=0$.
Given a closed $(p - 1)$-form $\sigma$ and closed $( d - p - 3)$-form  $\tilde{\sigma}$, one can construct the  charges
\begin{equation}
Q [\sigma] = \int_{  \Sigma} \sigma \wedge \ast j,
\qquad
Q [\tilde\sigma] = \int_{  \Sigma} \tilde\sigma \wedge \ast \tilde j
\label{qantia}
\end{equation}
by integrating over a $d-1$ dimensional space $\Sigma$, which will usually be chosen to be a hypersurface $\mathbb{R}^{d-1}$ of constant time, or a region of such a spatial hypersurface bounded by a closed $d-2$ dimensional surface $S$.
In the case in which $\Sigma$ is a hypersurface $\mathbb{R}^{d-1}$ of constant time, $S$ is the $d-2$ sphere at spatial infinity.

As the forms ${\sigma}$,$\tilde{\sigma}$ are closed and the currents $j,\tilde j$ are conserved, these charges defined on constant time surfaces are  independent of the time chosen and so are conserved, provided
 the currents $j,\tilde j$ vanish on $S$.\footnote{For conservation, it is sufficient that the total current  through the surface $S$ vanishes, $\int _S (\sigma \wedge \ast j)_i dS^{0i}=0$ where $(\sigma \wedge \ast j)_i$ is the spatial component of $\sigma \wedge \ast j$.}
Using (\ref{Ffida}), these  charges can be written as integrals over the $d-2$ dimensional boundary $S$ of $\Sigma$
\begin{equation}
Q [\sigma] = \int_{ S} \sigma \wedge \ast F,
\qquad
Q [\tilde\sigma] = \int_{  S} \tilde\sigma \wedge  F
\label{qantib}
\end{equation}
Note that these charges depend only on the cohomology classes of $\sigma, \tilde\sigma$.

The electric charges are   carried by electrically charged $p-1$ branes 
with $p$-form current $j$. 
 Then $Q[\sigma]$ is a charge for  $p-1$ branes wrapping the $p-1$ cycle dual to the cohomology class $\sigma$. For example, for a spacetime given by a product of a compact $p-1$ dimensional space $M$ and  $d-p+1$ dimensional Minkowski space, a  non-trivial charge arises by taking $\sigma$ to be the volume form of $M$, giving the charge of $p-1$ branes wrapping $M$.  For the case of a toroidal compactification $M=T^{p-1}$ and constant $\sigma$, $Q[\sigma]$ defines the standard $p-1$ form charge ${\cal Z}_{
\mu_1\dots\mu_{p_1}}$ by
 \begin{equation}
Q[\sigma]=\frac 1 {(p-1)!}\sigma_{
\mu_1\dots\mu_{p_1}}{\cal Z}^{
\mu_1\dots\mu_{p_1}}
\end{equation}
However, working with the charges $Q[\sigma]$ depending on cohomology classes is more useful here and includes cases in  which $\sigma $ is not constant.
 Similar remarks apply to the magnetic charges which are carried by $d-p-3$ branes.

If there are no magnetic sources,  i.e.\ if $\tilde j= 0$, 
 then  there is a local $p$-form potential $A $ with $F=dA$ and a gauge symmetry
  $\delta A = d \sigma$ with $(p - 1)$-form parameter $\sigma$.
 Any field configuration is {\it invariant} under a gauge transformation for which $\sigma$ is a closed form
satisfying
\begin{equation}d \sigma = 0 \end{equation}
The gauge parameters that are closed forms are the analogues of Killing vectors for gravity and can be viewed as representing a global symmetry of the system. 
For a given closed $p-1$ form $\sigma$, the
transformation $\delta A = d(\alpha \sigma)$ is an invariance for any constant 0-form parameter $\alpha$, while under this transformation for a {\emph {local}}  parameter $\alpha(x)$ the change in the action is proportional to   $\int d\alpha  \wedge \ast {\cal J}$
where $\ast {\cal J}=  \sigma \wedge \ast j$. Then, in this way, the one-form current ${\cal J}$ can be viewed as the Noether current  for the 0-form symmetry with parameter $\alpha$, with $Q[\sigma]= \int_{  \Sigma} \ast {\cal J}$ the corresponding Noether charge.

The  Noether charge $Q [\sigma] $   is the electric charge for the gauge theory of $A$.
More generally, if there are magnetic sources then $A$ can be introduced in the region in which $\tilde j= 0$
and the
surface integral form  (\ref{qantib})  of $Q [\sigma] $ remains defined  even if there are magnetic sources inside $S$.

Similarly, if 
there are no electric sources, i.e.\ if $  j= 0$, 
 then  there is a local $(  d - p - 2)$-form potential $\tilde A $ with $*F=d\tilde A$ and a gauge symmetry
  $\delta \tilde A = d \tilde \sigma$ with $(  d - p - 3)$-form parameter $\tilde\sigma$.
 A gauge transformation for which $\tilde\sigma$ is a closed form will leave any
  field configuration {\it invariant}.
  The  charge $Q  [\tilde\sigma] $ is the Noether charge associated with the global symmetry  $\delta \tilde A = d (\tilde \alpha \tilde \sigma)$ with 0-form parameter $\tilde \alpha$
  and is the electric charge for the dual gauge theory of $\tilde A$.
 Again, if there are electric sources the potential $\tilde A $  can be introduced in the region in which $  j= 0$ and
 the surface integral charge (\ref{qantib}) is
defined  even if $j \neq 0$ inside $S$.
However, the   charge $Q  [\tilde\sigma] $ given by the surface integral (\ref{qantib}) 
  can be interpreted as the magnetic charge for the  original gauge theory of the field $A$. 
  
  To understand the magnetic charge $Q[\tilde \sigma]$ in terms of the  theory of the gauge potential $A$, one must restrict to the subspace of spacetime on which $A$ is defined. 
 The field $A$ exists on the manifold given by removing from spacetime the regions in which   the magnetic source  $\tilde j$ is non-zero. The space given by cutting out the magnetic sources typically has non-trivial topology and $A$ is then a connection on a bundle over this space; the topology of this construction leads to the magnetic charge.
In this way,  a magnetic charge for the gauge theory  of $A$ can be constructed from the electric charge of the dual gauge theory.\footnote{Further charges can be defined by integrating the $(d-p)$-form current $\ast J$ over a $d -
p$-dimensional surface $S $
or the  $(p + 2)$-form current $\ast \tilde{J}$ over an $(p + 2)$-dimensional surface
$\tilde{S}$.
%\begin{equation}\int_S \ast J = \int_{  \partial S} \ast F, \qquad  \int_{\tilde{S}} \ast
%   \tilde{J} = \int_{  \partial \tilde{S}} F \end{equation}
There are also similar constructions
in which  $\sigma$ is
taken to be a
  closed $r$-form
  with $r < p - 1$ 
or $\tilde{\sigma}$ is taken to be
a  closed
$s$-form $s < D - p - 3$.}

The charges
$Q  [ \sigma] $, $Q  [\tilde\sigma] $ are   topological   in  the sense that $S$ 
 can be deformed to another surface $S'$ with $Q  [ \sigma] $ unchanged so long as no region with $j\ne 0$ is crossed in the deformation and
  $Q  [\tilde \sigma] $ unchanged so long as no region with $\tilde j\ne 0$ is crossed.

In this way, conserved electric and magnetic charges can be defined for the
theory with electric and magnetic sources using the surface integral charges (\ref{qantib}), giving a measure of the amount of electric and magnetic charge contained in $S$. In the absence of magnetic sources,
the electric charges can be viewed as generalised  Noether charges for the gauge theory of
$A$ while if there are no electric sources, the magnetic charges can be viewed
as generalised Noether charges for the dual gauge theory of $\tilde{A}$.

Consider now the case of linearised gravity, the theory of 
 a free 
symmetric tensor gauge field $h_{\mu\nu}$ in $d$-dimensional Minkowski space
with  
the gauge symmetry\footnote{Throughout this paper, Cartesian coordinates will be used with Minkowski metric $\eta _{\mu\nu}=diag(-1,1,1\dots1)$.}
\begin{equation} \delta h_{\mu\nu}=   \partial _{(\mu}\xi_{\nu)}
\label {hvara}
\end{equation}
It satisfies the linearised Einstein equations $G_{\mu\nu}=T_{\mu\nu}$ where $G_{\mu\nu}$ is
 the linearised Einstein tensor and $ T_{\mu\nu}$ is an energy-momentum tensor satisfying
 $\partial ^\mu T_{\mu\nu}=0$.
 Any field configuration is invariant under gauge transformations in which the parameter $\xi_{\mu}$ is a 
 Killing vector of the Minkowski background, i.e.\ the parameter  is a vector
  $k_\mu$ satisfying 
 \begin{equation}
  \partial_{(\mu}k_{\nu)}=0
\label{killlina}
\end{equation}
Then 
\begin{equation}j_\mu [k]= T_{\mu\nu}k^\nu
\label{jtis}
\end{equation}
is a conserved current 
 and so can be integrated over a constant time slice $\Sigma$ to give a conserved charge
\begin{equation}Q[k]= \int _\Sigma *j [k] 
\label{qkisa}
\end{equation}
Using the Einstein equations to write $j_\mu [k]= G_{\mu\nu}k^\nu$, the current can be written as the divergence of a 2-form current $J_{\mu\nu}[k]$ with $j_\mu [k]=\partial ^\nu J_{\mu\nu}[k]$. 
Here $J$ is a function of $h_{\mu\nu},k_\mu$ and their derivatives. Then 
the charge (\ref{qkisa}) can be rewritten as a surface integral
\begin{equation}Q[k]= \int _S *J [k] 
\label{qkisas}
\end{equation}
This is the  construction of \cite{Abbott:1981ff} of the  ADM momentum and angular momentum, restricted to linearised gravity, corresponding to the the Killing vectors generating the translations and Lorentz transformations of Minkowski space.

As for the $p$-form gauge fields, the invariance gives a 0-form symmetry for which the charge $Q[k]$ is the Noether charge.
The transformation (\ref{hvara}) with $\xi_\mu = \alpha k_\mu$
is an invariance for constant parameter $\alpha$ while 
for a local  parameter $\alpha(x)$ the change in the action is proportional to  
$\int j^\mu \partial _\mu \alpha$ so that the one-form current $j$ can be viewed as the Noether current associated with this 0-form symmetry. $Q[k]$ is the corresponding Noether charge and will be referred to as an electric-type charge for gravity.

Linearised gravity has a dual formulation in terms of a dual graviton  \cite{Hull:2000zn};
for further discussion of the dual graviton, see \cite{Hull:2001iu}-\cite{Bekaert:2003az}.
The next step   is to construct the electric-type (Noether) charges for the dual graviton and then seek to reinterpret them as magnetic-type charges for the graviton.
The structures in $d=4$ and $d>4$ dimensions are different and will be considered in turn.

In four dimensions,  the dual of the linearised curvature
$\tilde R=*R$ 
is
\begin{equation}\tilde R_{\mu\sigma 
\nu\rho }
=\frac 1 2
\epsilon _{\mu\sigma \alpha\beta}
R^{\alpha\beta }{}_{\nu\rho}
\label{sisasa}
\end{equation}
In the absence of sources, 
 $\tilde R_{\mu\sigma 
\nu\rho }
$ is the linearised curvature of a dual graviton $\tilde h_{\mu\nu}$ 
and linearised gravity has a dual formulation in terms of $\tilde h_{\mu\nu}$, with the duality interchanging field equations with Bianchi identities. One can introduce sources $ T_{\mu\nu}$ and $\tilde T_{\mu\nu}$
for the Einstein tensor and dual Einstein tensor
\begin{equation}
G_{\mu\nu}= T_{\mu\nu},
\qquad
\tilde G_{\mu\nu}=\tilde T_{\mu\nu}
\end{equation}
that generalise the electric and magnetic currents of electromagnetism.
The energy-momentum tensor $ T_{\mu\nu}$ is a source for  $\tilde R_{[\mu\sigma 
\nu]\rho }
$ so that in regions in which it is non-zero  $\tilde R_{\mu\sigma 
\nu\rho }
$ cannot be written in terms of a dual graviton $\tilde h_{\mu\nu}$ and similarly the magnetic source
$ \tilde T_{\mu\nu}$ is a source for  $ R_{[\mu\sigma 
\nu]\rho }
$ so that in regions in which it is non-zero  $ R_{\mu\sigma 
\nu\rho }
$ cannot be written in terms of a  graviton $ h_{\mu\nu}$. 

Charges can be constructed for the dual theory of $\tilde h_{\mu\nu}$ in the same way as they were for the original theory. 
If $T_{\mu\nu}=0$, the dual theory is the standard linearised gravity theory  for $\tilde h_{\mu\nu}$ with a source
$ \tilde T_{\mu\nu}$. Then for any Killing vector satisfying (\ref{killlina}) the current 
\begin{equation} 
\tilde j_\mu [k]=\tilde T_{\mu\nu}k^\nu
\label{jtist}
\end{equation}
is  conserved  
 and so can be integrated over a constant time slice $\Sigma$ to give a conserved charge
\begin{equation}\tilde Q[k]= \int _\Sigma * \tilde j [k] 
\label{qkistt}
\end{equation}
Using the dual Einstein equations to write $\tilde j_\mu [k]= \tilde G_{\mu\nu}k^\nu$, the current can be written as the divergence of a 2-form current $\tilde J_{\mu\nu}[k]$ with $\tilde j_\mu [k]=\partial ^\nu  \tilde J_{\mu\nu}[k]$. 
The current $\tilde J_{\mu\nu}[k]$ is obtained from $ J_{\mu\nu}[k]$ by replacing $h_{\mu\nu}$ with $ \tilde h_{\mu\nu}$.
Then 
the charge (\ref{qkistt}) can be rewritten as a surface integral
\begin{equation}\tilde Q[k]= \int _S *\tilde J [k] 
\label{qkisy}
\end{equation}
This is the Noether charge for the Killing vector invariance of the dual graviton theory.

If the surface $S$ is in a region in which both sources $ T_{\mu\nu},\tilde T_{\mu\nu}$ are zero, the theory near
$S$ can be written in terms of either $h$ or $\tilde h$ and the charge $\tilde Q[k]$ can be viewed as a magnetic charge for the linearised gravity theory formulated in terms of $h$.
In order to interpret the theory in terms of $h$, it is necessary to remove the regions in which there are non-trivial magnetic sources   $\tilde T_{\mu\nu}$, leaving a manifold with non-trivial topology that can support the magnetic charges.
As will be discussed in section \ref{4DSec}, each magnetic charge  written in terms of $h$ is the integral of a total derivative, indicating its topological nature.
%As will be seen in section  \ref{4DSec}, for t
For the translation Killing vectors the charges $ Q[k]$ can be written as
\begin{equation}
Q[k]=\int _S k^\mu  *\Gamma_\mu
\end{equation}
where $\Gamma_\mu =\frac 1 2 \Gamma_{\mu \nu\rho }dx^\nu \wedge dx^\rho$ is a certain connection 2-form constructed from $h$
while the magnetic charge is given by replacing $\Gamma_\mu\to *\Gamma_\mu$ in this expression:
\begin{equation}
\tilde Q[k]=\int _S k^\mu  \Gamma_\mu
\end{equation}
These charges for constant $k$   give the linearised version of the dual momentum or NUT 4-momentum  for general relativity introduced in \cite{Ramaswamy,Ashtekar}.
%The situation  for the Lorentz transformation Killing vectors will be discussed in section \ref{4DSec}.

The dual graviton theory is rather different in dimensions $d>4$.
It is a gauge field $D_{\mu_1\dots \mu_{d-3}\, \, \nu}$ of mixed symmetry, represented by a Young tableau with one
column of length $d-3$ and one of length $1$ \cite{Hull:2000zn}. 
Gauge fields of mixed symmetry were introduced by Curtright \cite{Curtright:1980yk}; see \cite{Hull:2001iu,deMedeiros:2002qpr,deMedeiros:2003osq,Bekaert:2002dt,Bekaert:2003az,Labastida:1986gy,Labastida:1986ft,Labastida:1987kw,Dubois-Violette:1999rd}
 for  further discussion of gauge fields of mixed symmetry. 
The dual of the curvature $R$ is
$S\equiv *R$, with components
\begin{equation}S_{\mu_{1}\mu_{2}\ldots\mu_{d-2}\;|\,
\nu\rho }
=\frac 1 2
\epsilon _{\mu_{1}\mu_{2}\ldots\mu_{d-2}\alpha\beta}
R^{\alpha\beta }{}_{\nu\rho}
\label{sist}
\end{equation}
Tensors represented by  a Young tableau with 2 columns
of lengths $p,q$ will be referred to as $[p,q]$ tensors or bi-forms, following
 \cite{deMedeiros:2002qpr,deMedeiros:2003osq}:
  a $[p,0]$ tensor is a $p$-form.\footnote{Here the notation $X_{\mu_1\ldots \mu_{p}  \, |\, \nu_1\ldots \nu_{q}}$ is used for the components of a $[p,q]$ bi-form $X$, with the two sets of antisymmetric indices separated by a $|$.} Then, if there are no sources, the curvature $R$ is a $[2,2]$ tensor  that is mapped by duality to a $[d-2,2]$ tensor $S$.
If there are no electric sources, $T=0$,  $S$ can be written in terms of a dual graviton field which is  a $[d-3,1]$ tensor $D$ with
\begin{equation}
S\sim {\cal Y}_{[d-2,2]}\partial \partial D
\end{equation}
where ${\cal Y}_{[p,q]} $ is the Young projector onto the $[p,q]$ representation.
As the gauge field $D$ has two  columns of unequal lengths (in $d>4$) there are {\it two} gauge invariances, one 
with a  gauge parameter that is a $[d-4,1]$ tensor $\alpha$ and the other which has
a  gauge parameter that is a 
  $d-3$ form   $\beta$. The gauge variation can be written as 
\begin{equation}
\delta D \sim {\cal Y}_{[d-3,1]} [\partial \alpha+\partial\beta]
\end{equation}
Note that the $\alpha$ symmetry is reducible (for $d> 5$) as  the transformation by \lq exact' bi-forms
$\alpha\sim {\cal Y}_{[d-4,1]} \partial \gamma$ do not act.

The invariances of a general field configuration arise when the gauge parameters are special tensors that generalise the Killing vectors that gave the invariances of linearised gravity.
A  $d-3$ form $\lambda$ satisfying
\begin{equation}
{\cal Y}_{[d-3,1]} \partial \lambda =0
\label{klam}
\end{equation}
is known as a Killing-Yano or Yano tensor and a gauge transformation with parameter
$\beta $ that is  a Killing-Yano tensor gives an invariance, $\delta D=0$. (Killing-Yano tensors and other generalised Killing tensors have been discussed in e.g. \cite{Howe:2015bdd,Howe:2018lwu} and references therein.)
For an $\alpha $ transformation to give an invariance, the parameter must be a $[d-4,1]$ tensor $\kappa$ satisfying
\begin{equation}
{\cal Y}_{[d-3,1]} \partial \kappa =0
\label{kkap}
\end{equation}
giving a new kind of generalised Killing tensor. The transformations by such Killing tensors $\lambda,\kappa$ are global symmetries and lead to Noether charges which are electric-type charges for the dual graviton theory.

As will be discussed in more detail in later sections, the field equation for the dual graviton $D$ is of the form
$E(D)=U$ where the generalisation of the Einstein tensor 
$E(D)$ is a $[d-3,1]$ tensor constructed from derivatives of $D$ that can be expressed
   in terms of traces of $S$, while $U$ is a conserved $[d-3,1]$ tensor that generalises the dual energy momentum tensor $\tilde T$ that arose in $d=4$. 
Then for each generalised Killing tensor $\kappa,\lambda$ there is a corresponding conserved current, given by
\begin{equation} 
j_\rho[\kappa  ]= \frac 1 {(n-1)!}  \kappa   ^{\mu_1\ldots \mu_{n-1}  \, |\, \nu}
U_{\rho \mu_{1}\mu_{2}\ldots\mu_{n-1}   \, |\, \nu }
, \qquad
j_\nu [\lambda ]= 
\frac 1 {n!} \lambda  ^ {\mu_1\ldots \mu_n} U_{ \mu_{1}\mu_{2}\ldots\mu_{n}   \, |\, \nu }
\label{llist}
\end{equation}
The conservation  of these currents follows from the generalised Killing conditions and the conservation of $U$.
Integrating these over $\Sigma$ gives conserved  charges $Q[\kappa]$ and $Q[\lambda]$ which are the Noether charges associated with the  symmetries of the dual graviton.
The currents (\ref{llist})  can be written in terms of $D$ using the dual Einstein equation $E=U$ and  can then be expressed as  divergences of  2-form currents
$J[\kappa  ]_{\mu\nu},
J[\lambda  ]_{\mu\nu}$: $j[\kappa  ]_\mu=\partial ^\nu J[\kappa  ]_{\mu\nu}$ and 
$j[\lambda  ]_\mu=\partial ^\nu J[\lambda  ]_{\mu\nu}$. Then the charges become surface integrals over the boundary $S$ of $\Sigma$,
\begin{equation}
Q[\kappa]=\int _S *J[\kappa  ] , \qquad Q[\lambda]=\int _S *J[\lambda  ] 
\end{equation}
The charge $Q[\kappa]$ only depends on a generalised cohomology of $\kappa$: shifting $\kappa$ by a certain kind of \lq exact' bi-form leaves the charge unchanged.

In regions in which $U=0$
 the dual theory can be formulated in terms of the graviton $h$   and the charges $Q[\kappa]$ and $Q[\lambda]$ can  be expressed in terms of $h$ instead of $D$. They can then be viewed as magnetic charges for linearised gravity. 
 The charges, when expressed in terms of $h$ are total derivatives, as is typical for magnetic charges. The charge $Q[\kappa]$ is
 \begin{equation}
Q[\kappa]=\int _S d * Z[\kappa]
\label {Qkax1}
\end{equation}
where $ * Z[\kappa]$ is a $d-3$ form with components
\begin{equation}
* Z[\kappa]_{\mu_{1} \ldots \mu_{d-3}}  =
(- )^{d}  \frac 1 {2}  \frac {d-3} {d-1}\, \, \,
h_{[\mu_1}{}^{\tau}
\kappa   _{\mu _2 \mu _2\ldots \mu _{d-3}]\,|  \tau}
\label {Qkax2}
\end{equation}
  while $Q[\lambda]$ for constant $\lambda$ is
  \begin{equation}
Q[\lambda]=\int _S d * Z[\lambda]
\label {Qkax3}
\end{equation}
where
  $ * Z[\lambda]$ is a $d-3$ form with components
  \begin{equation}
* Z_{\mu_{1} \ldots \mu_n}[\lambda] = %(-)^{d+1} 
\frac {d-3} {4(d-1)} \,
\, h^{\rho}{}_{[ \mu_{1}  }   \lambda  _ {\mu_{2} \ldots \mu_{d-3} ]\rho}
\label {Qkax4}
\end{equation}

The charges
  $Q[\kappa]$ and $Q[\lambda]$ are fully topological for the graviton theory in terms of $h$ as they are unchanged under any deformation of $S$. This is because the currents  
  $j_\mu [\kappa]=\partial ^\nu J_{\mu\nu}[\kappa]$
  and  
  $j_\mu [\lambda]=\partial ^\nu J_{\mu\nu}[\lambda]$
  are identically zero in the graviton theory as a result of  the gravitational Bianchi identity.
  % in the sense that each is unchanged under a deformation of $S$ that does not cross a region in which the corresponding current $j[k]$, $j[\kappa]$ or $j[\lambda]$ is non-zero.

The magnetic charge (\ref{qantib}) for the $p$-form gauge theory  is a total derivative
\begin{equation}
Q [\tilde\sigma] = \int_{  S} d(\tilde\sigma \wedge  A)
\label{qantiba}
\end{equation}
If $A$ is a regular, globally-defined $p$-form, then this charge  vanishes. Non-zero magnetic charge only arises if $A$ has a Dirac string singularity, or is defined locally in patches 
on a space given by removing a  region or regions from Minkowski space,  with transition functions involving  the gauge transformation
$A\to A+d\lambda$. 
Removing the regions  can leave a space with non-trivial topology that can support a non-trivial bundle; the removed regions are can be associated with the locations of magnetic sources.
Similarly, if $h$ is a globally defined tensor, then the gravitational magnetic charges (\ref{Qkax1}) and (\ref{Qkax3}) vanish as they are integrals of exact forms. To obtain non-zero magnetic charges, it is necessary that 
$h$ has Dirac string singularities, or is defined locally in patches on Minkowski space with some regions removed, with transition functions involving  the gauge transformation (\ref{hvara}). This is discussed further in subsection \ref{topy} and examples of solutions with non-zero gravitational magnetic charges are given in section \ref{sols}.

Similar results follow for other free gauge theories; the extension to interacting theories will be considered elsewhere. The gauge transformations that leave any configuration invariant have parameters that satisfy a Killing condition generalising the Killing vector condition (\ref{killlina}) and the Killing tensor conditions (\ref{klam}),(\ref{kkap}) and for each such Killing parameter there is a  Noether current $j_\mu$
(for the 0-form symmetry in which the gauge parameter is given by a 0-form parameter times a Killing tensor).
 The current is conserved, i.e.\ the 1-form $j$ is co-closed,
$$d*j=0$$
and can be integrated to give a conserved charge
$$
Q=\int_\Sigma *j
$$
For example, consider a symmetric tensor gauge field $H_{\mu_1\dots \mu_s}=H_{(\mu_1\dots \mu_s)}$ with
gauge symmetry $\delta H_{\mu_1\dots \mu_s}=\partial_{(\mu_1}\xi_{\mu_2\dots \mu_s)}$. The invariances are gauge transformations with parameters $\xi_{\mu_1\dots \mu_{s-1}}$ that are Killing tensors on Minkowski space, i.e.\ symmetric tensors $K_{\mu_1\dots \mu_{s-1}}$ satisfying 
$$
 \partial_{(\mu_1}K_{\mu_2\dots \mu_s)}=0$$
The gauge field couples to a  source $\Theta _{\mu_1\dots \mu_s}=\Theta_{(\mu_1\dots \mu_s)}$ that is conserved,
$\partial ^{\mu_1}\Theta_{\mu_1\dots \mu_s}=0$. Then the following Noether current is conserved:
$$j_\nu=K^{\mu_1\dots \mu_{s-1}}\Theta_{\nu \mu_1\dots \mu_{s-1}}
$$
This can then be re-expressed as a surface integral of an integrand that depends on $H$ and $K$.

For any free gauge theory, the 1-form Noether current $j$ is conserved. There is  an associated 2-form current $J$ with
\begin{equation}
*j=d*J
\label{jcocl}
\end{equation}
so that $J$ is a conserved 2-form current in any region in which $j=0$, which will be referred to here as a {\it secondary Noether current}. This then can be integrated to give a charge 
$$Q'=\int_S *J
$$
and $Q=Q'$ if $S$ is the boundary of $\Sigma$.
In the Euclidean theory, $S$ can be any closed $d-2$ surface and $\Sigma$ can be taken to be any $d-1$ dimensional subspace with boundary $S$. If $\Sigma$ has no boundary, the charge will vanish if $*j$ is exact, so that $J$ is globally well-defined.
The charge defined by $\int_\Sigma *j$ for a   $d-1$ surface  $\Sigma$ is topological in the sense that it is invariant under any smooth deformation of $\Sigma$.
Similarly,
the charge defined by 
$\int_S *J$ for a  $d-2$ surface $S$ is topological in the sense that it is invariant under any smooth deformation of   $S$
that does not involve passing through a region in which $j\ne 0$.

For a given current $j$, (\ref{jcocl}) only determines $J$ up to the addition of a co-closed form. For any $J$ satisfying (\ref{jcocl}), further solutions are given by $J'=J+d^\dagger K$   for any 3-form $K$, where $d^\dagger=*d*$ and $J'=J+L$ where $d^\dagger L=0$.

The current $j$ corresponds to the charge density for the source of the gauge field -- for electromagnetism, $j_0[\sigma]$ is the electric charge density, $j_0[\tilde \sigma]$ is the magnetic charge density and for gravity (with $k^\mu=\delta^\mu _0$ the timelike Killing vector) 
$j_0[k]=T_{00}$ is the mass density. If $\Sigma $ is chosen to be a region of space at fixed time, then $\int_\Sigma *j$ gives the total charge in that region and this can be re-expressed as $\int_S *J$ where $S$ is the boundary of $\Sigma $.
The regions in which $j=0$ are then regions in which there is no local charge. For localised systems 
 all the charge is confined to  some region   of spacetime, and outside this region $j=0$.

On general grounds \cite{Lee:1990nz,Wald:1999wa,Barnich:1994db,Barnich:2001jy,Barnich:2000zw}, the 2-form current $J$ has a local expression in terms of the gauge field and the Killing tensor. While the Noether current $j$ is typically gauge invariant, the secondary Noether current $J$ is often not, changing by a co-exact term
$$\delta *J= d \Delta$$
for some $ \Delta$.
Then in such cases $*J$ can be regarded as a $d-2$ form gauge field $B=*J$ with gauge transformation
$\delta B= d \Delta$ and which is flat, $dB=0$, in regions in which $j=0$. Then, for the charge $Q'$, the operator
$\exp \left(iQ'\right)$ is associated with the surface operator $\exp \left(i\int_S B \right)$.

In this way, the integrals of $j$ and $J$ define \lq electric' charges of a given gauge theory.
In any region in which $j=0$ so that $J$ is conserved, the free gauge theory has  a dual formulation in terms of a dual gauge field. Then $J$ can be regarded as a current in the dual theory whose integral can be viewed as a 
\lq magnetic' charge of the dual theory.
As will be seen, in  some cases the current $J$ can be written as a local expression in terms of the dual fields and in some cases it can't.
This will be applied to the dual graviton theory here: the gauge symmetries of the dual graviton give  Noether currents $j$  and secondary currents $J$. Then, in regions without sources for the dual graviton, the 2-form current $J$  can be regarded  as a conserved current for the dual of the dual graviton theory, which is the theory of the graviton $h_{\mu \nu}$. Then the integral of these currents give magnetic charges for the graviton.

\section{Linearised Gravity and Dual Gravity}

\subsection{Linearised Gravity}

The linearised graviton in $d$-dimensional Minkowski space with a flat background metric 
$\eta_{\mu\nu}$ (which is used to raise and lower indices) is a free 
symmetric tensor gauge field $h_{\mu\nu}$ 
which has  
the gauge symmetry 
\begin{equation} \delta h_{\mu\nu}=   \partial _{(\mu}\xi_{\nu)}
\label {hvar}
\end{equation}
The invariant field strength is the linearised Riemann tensor\footnote{The square brackets on indices denotes antisymmetrisation with strength one, so that e.g. $T_{[\mu\nu]}=\frac 1 2 (T_{\mu\nu}-T_{\nu\mu})$.}
\begin{equation}
R_{\mu\nu\, \sigma\tau}= \frac 1 2 ( \partial_{\mu}   \partial_{\sigma}h_{\nu\tau}+\ldots)=-2  \partial_{[\mu}h_{\nu][\sigma,\tau]}
\label {ris}
\end{equation}
which  satisfies 
\begin{equation}R_{\mu\nu\, \sigma\tau}=R_{\sigma\tau\,\mu\nu}
\label {rtrans}
\end{equation}
together with
the first Bianchi identity
\begin{equation}R_{[\mu\nu\, \sigma]\tau} =0
\label{rbo}
\end{equation}
and the second Bianchi identity
\begin{equation}
R_{\mu\nu\, [\rho \sigma ,\tau ]}=0
\label{rbt}
\end{equation}
The  free field equation  in $d\ge 4$ is the linearised 
Einstein equation
\begin{equation}R _{\mu\, \nu}=0
\label{rfo}
\end{equation}
where the linearised Ricci tensor is 
\begin{equation}R _{\mu\, \nu}=R^\sigma{}_{\mu\, \sigma\nu}
\label{ricis}
\end{equation}
This, together with the second Bianchi identity (\ref{rbt}), %\rbo\ 
 implies 
\begin{equation}  \partial ^\mu R_{\mu\nu\, \sigma\tau}=0
\label{rft}
\end{equation}
where indices are raised and lowered with the flat background metric 
$\eta_{\mu\nu}$.

It will be useful to introduce a linearised connection
 \begin{equation}
 \Gamma _{ \mu\nu  \, \tau }=   \partial _{[\mu }h_{\nu] \tau}
 \label{conne}
 \end{equation}
in terms of which the curvature is
\begin{equation}
 R^{\mu\nu}{}_{ \sigma\tau}
=
  \partial_{\sigma }\Gamma ^ {\mu\nu}  {}_{\tau} 
- \partial_{ \tau}\Gamma ^ {\mu\nu}{}_{\sigma } 
\label {rcon}
\end{equation}
This formula for the curvature is invariant under
\begin{equation}
 \Gamma _{ \mu\nu \, \tau }
 \to
  \Gamma'  _{ \mu\nu \, \tau }
= \Gamma _{ \mu\nu \, \tau }
+   \partial_{\tau }V_{\mu\nu}
  \end{equation}
for any $V_{\mu\nu}$, which need not be antisymmetric. In particular, choosing $V_{\mu\nu}=\frac 1 2 h_{\mu\nu}$
gives a $\Gamma'  _{ \mu\nu \, \tau }
$ which is the linearised Christoffel connection
\begin{equation}
C_{\nu \, \tau  \mu}= \frac 1 2 \left(  \partial_ \mu h_{\nu \tau}+  \partial_{ \tau} h_{\mu\nu}
-  \partial_{ \nu} h_{\mu\tau}
\right)
\end{equation}
Note that the torsion given by the antisymmetric part of the connection  vanishes:
\begin{equation}
T _{ \mu\nu \tau }\equiv \Gamma _{[ \mu\nu \, \tau] }=0
\label{toris}
\end{equation}
The connection also satisfies
\begin{equation}
 \partial _{[\rho} \Gamma _{ \mu\nu] \, \tau } = 0
 \end{equation}

%%%
\subsection{Global Structure and Topological Charges} 
\label{topy}

Gravity linearised around a background spacetime with background metric $\bar{g}_{\mu \nu}$ is obtained
from the non-linear theory by writing the metric as $g_{\mu
\nu}=\bar{g}_{\mu \nu}+h_{\mu \nu}$ and linearising in the fluctuation
$h_{\mu \nu}$. The theory is invariant under diffeomorphisms under which
$x^{\mu} {\rightarrow x'}^{\mu}$ and tensor gauge transformations with finite one-form
parameter $\xi_{\mu}$ with the fields transforming as
\begin{equation}
  \bar{g}_{\mu \nu} \rightarrow \bar{g}_{\mu \nu}', \quad h_{\mu \nu}
  \rightarrow h_{\mu \nu}' + 2 \bar{\nabla}_{(\mu } \xi_{
  \nu)}
  \label {assdfaq}
\end{equation}
where
\begin{equation}
  \bar{g}'_{\mu \nu} (x') = \bar{g}_{\rho \sigma} (x) \frac{\partial
  x^{\rho}}{{\partial x'}^{\mu}}  \frac{\partial x^{\sigma}}{{\partial
  x'}^{\nu}} \label{transf}, \qquad h'_{\mu \nu} (x') = h_{\rho \sigma} (x)
  \frac{\partial x^{\rho}}{{\partial x'}^{\mu}}  \frac{\partial
  x^{\sigma}}{{\partial x'}^{\nu}}
\end{equation}
For infinitesimal diffeomorphisms under which $x^{\mu} {\rightarrow x'}^{\mu}
= x^{\mu} - \zeta^{\mu}$ with infinitesimal parameter $\zeta_{\mu}$ these
become
\begin{equation}
  \delta \bar{g}_{\mu \nu} =\mathcal{L}_{\zeta}  \bar{g}_{\mu \nu} = 2
  \bar{\nabla}_{(\mu } \zeta_{ \nu)}, \quad \delta h_{\mu
  \nu} =\mathcal{L}_{\zeta} h_{\mu \nu} + 2 \bar{\nabla}_{(\mu }
  \xi_{ \nu)}
\end{equation}
Similarly, a 1-form gauge field $A_{\mu}$ on same the background spacetime transforms under diffeomorphisms and
gauge transformations as
\begin{equation}
  A_{\mu} \rightarrow A'_{\mu} + \partial_{\mu} \lambda
   \label {assdfaq2}
\end{equation}
with
\begin{equation}
  A'_{\mu} (x') = A_{\rho} (x) \frac{\partial x^{\rho}}{{\partial x'}^{\mu}} 
\end{equation}
The background spacetime will in general be covered by a set of coordinate
patches with fields defined in each patch. In the overlaps between coordinate
patches the fields $A_{\mu}, \bar{g}_{\mu \nu}, h_{\mu \nu}$ in the two
patches are related by transition functions that are diffeomophisms combined
with gauge transformations of the form (\ref{assdfaq}),(\ref{assdfaq2}).

If the transition functions for a particular configuration of the field
$A_{\mu}$ are all just diffeomorphisms with no gauge transformations, then
$A_{\mu}$ are the components of a globally defined 1-form and the gauge bundle
is trivial. In particular, the magnetic charge $\int d A$, or  (\ref{qantib}) which gives
$Q [\tilde\sigma] = \int_{  S} d( \tilde\sigma \wedge  A)$,
 is then a
total derivative of a globally defined form and so is zero when integrated
over a closed surface. The magnetic charge can only be non-zero if the transition
functions involve gauge transformations and $A_{\mu}$ is a connection on a
non-trivial bundle. 

The situation is similar for $h_{\mu \nu}$. If the transition functions for a
particular configuration of the graviton field $h_{\mu \nu}$ are all just
diffeomorphisms with no gauge transformations, then $h_{\mu \nu}$ are the
components of a globally defined tensor and magnetic charges such as (\ref{Qkax1}) or (\ref{Qkax3})
are zero. To obtain non-zero magnetic charges, it is necessary that $h_{\mu
\nu}$ is not a globally defined tensor and has transition functions with
non-trivial gauge transformations. See \cite{Hull:2023dgp} for further discussion of transition functions for symmetric tensor gauge fields.

In this paper, the background is taken to be Minkowski space, or Minkowski
space with a region or regions removed, with a globally defined Cartesian coordinate
system and metric $\bar{g}_{\mu \nu} = \eta_{\mu \nu}$. Then there are no
diffeomorphisms in the transition functions (as the coordinates are defined over the whole space) and magnetic charges require that
the gauge fields $A_{\mu}$ or $h_{\mu \nu}$ are defined in patches with
non-trivial gauge transformations relating the gauge fields in different
patches. Alternatively, if one tried to define a gauge field $A_{\mu}$ or
$h_{\mu \nu}$ with magnetic charge globally in Minkowski space (with regions removed)
without
introducing different patches, then the gauge field would  have a Dirac
string singularity. The removed regions are associated with the locations of magnetic sources.

\subsection{Young Tableaux}

The representation of $GL(d,\mathbb{R})$ corresponding to the Young tableau with columns of length $p,q, \dots ,r$ where $p\ge q\ge  \dots \ge r$
will be referred to here as  the $[p,q,\dots ,r] $ representation, following \cite{deMedeiros:2002qpr,deMedeiros:2003osq}.
Then
$[n,1]$ corresponds to the
\lq hook' Young tableau with one column of length $n$ and one of length $1$, so that   e.g. for $n=4$ the Young tableau for the  $[4,1]$ representation is
$$ D: \tableau{2 1 1 1}$$
A tensor in the $[p]$ representation is a $p$-form.
A tensor in the $[p,q]$ representation has components
$T_{\mu_{1}\mu_{2}\ldots\mu_{p}\,  |\, \nu_{1}\nu_{2}\ldots\nu_{q}}$ where a $|$ will usually be inserted to separate the two antisymmetric groups of indices.
Such tensors are sometimes referred to as biforms \cite{deMedeiros:2002qpr,Bekaert:2002dt}  and satisfy
\begin{equation} 
T_{\mu_{1}\mu_{2}\ldots\mu_{p}\,  |\nu_{1}\nu_{2}\ldots\nu_{q}}
=T_{[\mu_{1}\mu_{2}\ldots\mu_{p}]\,  |\, \nu_{1}\nu_{2}\ldots\nu_{q}}=T_{\mu_{1}\mu_{2}\ldots\mu_{p}\,  |\, [\nu_{1}\nu_{2}\ldots\nu_{q}]}
\end{equation}
together with
\begin{equation}
T_{[\mu_{1}\mu_{2}\ldots\mu_{p}\,  |\nu_{1}]\nu_{2}\ldots\nu_{q}}=0
\end{equation}
%where  the square brackets in $[\mu_{1}\mu_{2}\ldots\mu_{p}]$ indicate antisymmetrisation with strength one.
If $p=q$, the tensor must in addition satisfy
\begin{equation} 
T_{\mu_{1}\mu_{2}\ldots\mu_{p}\,  |\nu_{1}\nu_{2}\ldots\nu_{p}}
=
T_{\nu_{1}\nu_{2}\ldots\nu_{p}
\,  |\mu_{1}\mu_{2}\ldots\mu_{p}
}
\end{equation}

It will be useful to use the notation  
\begin{equation}
\delta ^{\mu_{1}\mu_{2}\ldots\mu_{r} } _ {\rho_{1}\rho_{2}\ldots \rho_{r} } =\delta ^{ [\mu_{1}  }
_ {\rho_{1}}
\delta ^{\mu_{2}} _ {\rho_{2}} \ldots \delta^{\mu_{r}] } _ {\rho_{r}}
\end{equation}
and define the tensor given by  raising the lower indices on this:
\begin{equation}
\eta ^{\mu_{1}\mu_{2}\ldots\mu_{r} | \nu_{1}\nu_{2}\ldots \nu_{r} } =\eta ^{\nu_1 \rho_1}\eta ^{\nu_2 \rho_2}\dots \eta ^{\nu_r \rho_r}
\delta ^{\mu_{1}\mu_{2}\ldots\mu_{r} } _ {\rho_{1}\rho_{2}\ldots \rho_{r} } 
\end{equation}
This will facilitate the writing of simple forms for the Einstein tensors for gravitons and dual gravitons.
The signature will be taken to be Lorentzian so that the following identity holds
\begin{eqnarray}
\epsilon^{\alpha_1 \ldots \alpha_r \alpha_{r+1} \ldots \alpha_d} \epsilon_{\alpha_1 \ldots \alpha_r \beta_{r+1} \ldots \beta_d} & = & - r! (d-r)!
 \; \delta^{[ \alpha_{r+1}}_{~\beta_{1+r}}\delta^{ \alpha_{r+2}}_{~\beta_{r+2}} \ldots \delta^{\alpha_d]}_{~\beta_d} 
 %\nonumber
 \label{epep}
\end{eqnarray}
and for an $r$-form $\omega \in \Omega^r$
the Hodge dual 
has components  $\star \omega $ 
\begin{eqnarray}
(\star \; \omega) _{\mu_{r+1} \ldots \mu_d}  = \frac{ 1 }{ r!  } \omega_{\mu_1 \ldots \mu_r} \epsilon^{\mu_1 \ldots \mu_r}_{~~~~~~\mu_{r+1} \ldots \mu_d} 
\end{eqnarray}
and the Hodge dual
satisfies
  \begin{eqnarray}
  \star \star \; \omega & = & - (-1)^{r (d-r)} \omega
\end{eqnarray}

\subsection{The Dual Graviton}

In this section, the dual formulation of linearised gravity in $d$ dimensional Minkowski space in terms of a dual graviton field
will be reviewed, following   the treatment in \cite{Hull:2000zn,Hull:2001iu}.
The dual graviton in $d$ dimensions is a 
gauge field  $D_{\mu_{1}\mu_{2}\ldots\mu_{n}\, |\, \nu}$ with $n+1$ indices
where
\begin{equation}n=d-3
 \end{equation} 
 satisfying
\begin{equation}D_{\mu_{1}\mu_{2}\ldots\mu_{n}\, |\,\nu}
=D_{[\mu_{1}\mu_{2}\ldots\mu_{n}]\, |\,\nu}
 \end{equation}
and
\begin{equation}D_{[\mu_{1}\mu_{2}\ldots\mu_{n}\, |\,\nu]}
=0 \end{equation}
so that it has mixed symmetry corresponding to the $[n,1]$ representation of $GL(d,\mathbb{R})$,
 e.g. for $n=4, d=7$ the Young tableau is
$$ D: \tableau{2 1 1 1}$$
 In dimensions $d\ge 5$ it has two   gauge transformations, one with parameter $\alpha$ in the $[n-1,1]$ 
representation 
\begin{equation}
\alpha _{\mu_1\ldots \mu_{n-1} \, |\,\rho}= \alpha _{[\mu_1\ldots \mu_{n-1}] \,|\, \rho}, \qquad
\alpha _{[\mu_1\ldots \mu_{n-1} \, |\,\rho]}=0,\end{equation}
and one with an $n$-form parameter $\beta$ in the $[n]$ representation 
\begin{equation}\beta _{\mu_1\ldots \mu_n}= \beta _{[\mu_1\ldots \mu_n]}
 \end{equation}
The corresponding  Young tableaux for $n=4$ are:
$$ \alpha: \tableau{2 1 1} \qquad \beta: \tableau{1 1 1 1}$$
The gauge variations are
\begin{eqnarray}
 \delta
D_{\mu \nu \ldots \sigma \, |\,\rho}
&=&   \partial _{[\mu} \alpha _{\nu\ldots \sigma] \,|\,\rho} 
+
    \partial _ \rho \beta _{\mu \nu \ldots \sigma}-  \partial _ {[\rho} \beta _{\mu \nu
\ldots
\sigma]}
%\\
\label{gaugvar}
\end{eqnarray}
The derivatives of the parameters $  \partial _\rho \alpha _{\mu_1\ldots \mu_{n-1} \, \nu}$ and 
$  \partial _\rho  \beta _{\mu_1\ldots \mu_n}$
are in the $$[n-1,1]\times [1]= [n,1]+[n-1,2]+ [n-1,1,1]$$ and
$$[n]\times [1]=[n+1]+[n,1]$$ representations respectively, and the gauge variation (\ref{gaugvar}) is the projection of these  derivatives to  the $[n,1]$ part.
The $\alpha $ gauge symmetry is reducible: the gauge variation vanishes for
parameters $\alpha $  that are \lq exact', satisfying
\begin{equation}
\alpha _{\mu_1\ldots \mu_{n-1} \,|\, \rho}=
   \partial _{[\mu_1}
\gamma _{\mu_2\ldots \mu_{n-1}] \, |\,\rho}
\end{equation}
for some $ \gamma _{\mu_1\ldots \mu_{n-2} \,|\, \rho}$ in the $[n-2,1]$ representation.

 The invariant field strength is an $[n+1,2]$ tensor given by 
\begin{equation}S_{\mu\nu\ldots \rho\;|\, \sigma\tau} =  \partial _{[\mu} D_{\nu\ldots \rho]\, |\, [\sigma,\tau]}
\label {dufistr}
 \end{equation}
so that for $n=4$ the tableau is
$$S: \tableau{2 2 1 1 1}$$
It satisfies the first Bianchi identities
\begin{equation}S_{[\mu_{1}\mu_{2}\ldots\mu_{n+1}\,|\,
\nu]\rho }=0, \qquad 
S_{\sigma[\mu_{1}\mu_{2}\ldots\mu_{n}\,|\,
\nu  \rho]}=
0
\label{sbo}
\end{equation}
 and the second
Bianchi identities
\begin{equation}
  \partial_{[\sigma}S_{\mu_{1}\mu_{2}\ldots\mu_{n+1}]\,|\,
\nu\rho }=
0,
\qquad
S_{\mu_{1}\mu_{2}\ldots\mu_{n+1}\,|\,
[\nu  \rho,\sigma]}=
0
\label{sft}
\end{equation}

The natural free field equation  in $d\ge 4$ is the dual 
Einstein equation 
\begin{equation}S' _{\mu_{1}\mu_{2}\ldots\mu_{n}\,|\,
\nu }=
0\end{equation}
where  $S'$ is the dual Ricci tensor, which is the $[n,1]$ tensor given by one contraction of the field strength
\begin{equation}S'_{\mu_{1}\mu_{2}\ldots\mu_{n}\,|\,
\nu }= S_{\mu_{1}\mu_{2}\ldots\mu_{n}\rho\,|\,
\nu }{}^{\rho}
\label{sfo}
\end{equation}
The double-trace gives an $(n-1)$-form
\begin{equation}
S''_{\mu_{1}\mu_{2}\ldots\mu_{n-1}}=
S_{\mu_{1}\mu_{2}\ldots\mu_{n-1}\nu \rho\, |\, }{}^{\nu \rho}
\label {sfoo}
\end{equation}
The field equation $S'=0$, together with the second Bianchi identity (\ref{rbt}), %\rbo\ 
 implies 
\begin{equation}
  \partial^{\sigma} S_{\sigma\mu_{1}\mu_{2}\ldots\mu_{n}\,|\,
\nu\rho }=
0,  
 \qquad   \partial^{\rho}
S_{\mu_{1}\mu_{2}\ldots\mu_{n+1}\,|\,
\nu\rho }=0
\end{equation}

There are two kinds of dual connection that can be introduced for the dual graviton.
These are
\begin{equation}
 \tilde \Gamma _{\mu_{1}\mu_{2}\ldots\mu_{n+1}
 \, 
 \nu
  }=   \partial _{[\mu_1}
 D_{\mu_{2}\ldots\mu_{n+1}]\, \nu}
 \label{duconn}
 \end{equation}
and
\begin{equation}
 \hat  \Gamma _{\mu_{1}\mu_{3}\ldots\mu_{n}} {}
_{\sigma\tau }=
D_{\mu_{1}\mu_{3}\ldots\mu_{n}} {}
_{[ \sigma, \tau ] }
\label{hatducon}
\end{equation}
in terms of which the field strength can be written as 
\begin{equation}
S_{\mu_{1}\mu_{2}\ldots\mu_{n+1}} {}
_{\nu \sigma }
=
 \tilde \Gamma _{\mu_{1}\mu_{2}\ldots\mu_{n+1}  \,[ \nu, \sigma] }
 \label {sconta}
\end{equation}
or
 \begin{equation}
S_{\mu_{1}\mu_{2}\ldots\mu_{n+1}} {}
^{\sigma\tau }
=
%(n+1)  
 \partial_{[\mu_{1} } \hat  \Gamma _{\mu_{2}\mu_{3}\ldots\mu_{n+1}]} {}
^{\sigma\tau }
\label {scont}
\end{equation}

\subsection{Gravitational Duality}

Gravitational duality \cite{Hull:2000zn,Hull:2001iu} results from identifying $S(D)$ with the dual of $R(h)$,
$S=*R$ 
so that
\begin{equation}S_{\mu_{1}\mu_{2}\ldots\mu_{n+1}\;|\,
\nu\rho }
=\frac 1 2
\epsilon _{\mu_{1}\mu_{2}\ldots\mu_{n+1}\alpha\beta}
R^{\alpha\beta }{}_{\nu\rho}
\label {sis}
\end{equation}
Then the graviton theory and the dual graviton theory give dual equivalent descriptions of the same system (in the absence of sources), with field equations of one formulation becoming the Bianchi identities of the other.
The field equation $R _{\mu\, \nu}=0$ for $h$ becomes the first  Bianchi identity for $D$, which is
$S_{[\mu_{1}\mu_{2}\ldots\mu_{n+1}\, 
\nu]\rho }=0$, while the first Bianchi identity for $h$, $R_{[\mu\nu\, \sigma]\tau} =0$, becomes the field equation for $D$, which is $S'_{\mu_{1}\mu_{2}\ldots\mu_{n}\,
\nu }=
0$.
Furthermore,
$  \partial _{[\rho}R_{\mu\nu]\, \sigma\tau}=0$ becomes
\begin{equation}  \partial^{\sigma} S_{\sigma\mu_{1}\mu_{2}\ldots\mu_{n}\,|\,
\nu\rho }=
0,  \qquad
S_{\mu_{1}\mu_{2}\ldots\mu_{n+1}\,|\,
[\nu  \rho,\sigma]}=
0
\label{sfto}
\end{equation}
while $  \partial ^\mu R_{\mu\nu\, \sigma\tau}=0
$ becomes
\begin{equation}  \partial_{[\sigma}S_{\mu_{1}\mu_{2}\ldots\mu_{n+1}]\,|\,
\nu\rho }=
0,  \qquad   \partial^{\rho}
S_{\mu_{1}\mu_{2}\ldots\mu_{n+1}\,|\,
\nu\rho }=0
\label{sbto}
\end{equation}

The fields $h$ and $D$ are non-locally related by the duality constraint, although the curvatures $R,S$ are locally related by (\ref{sis}). This then implies that, up to gauge transformations, the connections $\Gamma(h)$ given by (\ref{conne}) and $\tilde \Gamma (D)$ given  by (\ref{duconn}) can be taken to be dual,
\begin{equation}
  \Gamma _{ \nu\gamma  \, \lambda }(h)
=
-\frac 1 2
(*\tilde \Gamma ) _{ \nu\gamma  \, \lambda } (D)
\label{ducon}
\end{equation}
where  $(*\tilde \Gamma ) _{\tau} {}^ {\mu\nu}(D)$ is  the dual of the connection $\tilde \Gamma$   given by
\begin{equation}
\tilde \Gamma ^ {\mu_{1}\mu_{2}\ldots\mu_{n+1}} {}_{\tau} 
= \frac 1 2 \epsilon ^{\mu_{1}\mu_{2}\ldots\mu_{n+1} }{}_{\mu\nu}\,  (*\tilde \Gamma )  ^ {\mu\nu}{}_{\tau} 
\label{ducond}
\end{equation}

Choosing the connections to satisfy the duality condition (\ref{ducond}) amounts to imposing a gauge condition, as will now be shown.
The dual connection $\tilde \Gamma (D)$ defined by  (\ref{duconn}) satisfies
\begin{equation}
 \tilde \Gamma _{[\mu_{1}\mu_{2}\ldots\mu_{n+1}
 \, 
 \nu]
  }=  0
 \label{duconnanti}
 \end{equation}
 so that (\ref{ducon}) implies that $\Gamma(h)$   is traceless
 \begin{equation}
  \Gamma _{ \nu\gamma   }{}^\nu(h)=0\label{duconsdfs}
\end{equation}
which implies the gauge condition
 \begin{equation}
    \partial _\mu 
  (h^{\mu\nu}-\eta ^{\mu\nu}h)=0
  \label{gaga}
\end{equation}
Thus  the duality condition (\ref{ducon}) implies the gauge choice (\ref{gaga}). Similarly, $\Gamma _{ [\nu\gamma  \, \lambda ]}(h)=0$ implies that 
$\tilde \Gamma(D)$ is traceless
\begin{equation}
\tilde \Gamma ^ {\mu_{1}\mu_{2}\ldots\mu_{n+1}} {}_{\mu_{1}}
= 0\label{ducondsadfas}
\end{equation}
giving a similar gauge condition for $D$.

\section{Sources for Linearised Gravity and Dual Gravity}

For linearised gravity, the key to coupling to a source is the 
existence of the Einstein tensor
\begin{equation}G_{\mu\nu}=R_{\mu\nu}- \frac 1 2 \eta _{\mu\nu} R
\end{equation}
which is identically conserved
\begin{equation}  \partial ^\mu G_{\mu\nu}=0
\label{cbi}\end{equation}
as a result of the Bianchi identities.
This can be seen by writing
\begin{equation}G_\mu{}^\alpha=- \frac  3 2
\delta _{\mu\nu\rho}^{\alpha\beta\gamma} \, R^  {\nu\rho} {}_  {\beta\gamma}
\label{eindel}
\end{equation}
so that (\ref{cbi}) immediately follows from (\ref{rbt}).
The graviton can be consistently coupled to an energy-momentum tensor source
$T_{\mu\nu}$ through the linearised Einstein equation
 \begin{equation}G_{\mu\nu}=T_{\mu\nu}
 \label {eins}
 \end{equation}
 provided   $T_{\mu\nu}$ is conserved
\begin{equation}  \partial ^\mu T_{\mu\nu}=0
\label{tcon} \end{equation}
The corresponding linearised Einstein-Hilbert action can be written as \cite{deMedeiros:2002qpr}
\begin{equation}
S=\int d^dx \left(
 \frac 1 2 h^{\mu\nu} G_{\mu\nu}-h^{\mu\nu}T_{\mu\nu}
\right)
\end{equation}
and is  invariant under the gauge transformation (\ref{hvar}) as a result of (\ref{cbi}) and (\ref{tcon}).

For dual gravity, there is an Einstein-like tensor \cite{deMedeiros:2002qpr}
\begin{equation}E_{\mu_{1}\mu_{2}\ldots\mu_{n}\,|\,
\nu }=- \frac 2 {n+2} 
\left(
S'_{\mu_{1}\mu_{2}\ldots\mu_{n}\,|\,
\nu }-
\frac n 2
S'' _{[\mu_{1}\mu_{2}\ldots \mu_{n-1}}
\eta_{ \mu_{n}]
\nu }
\right)
\end{equation}
(with $S',S''$ defined in (\ref{sfo}),(\ref{sfoo}))
which is identically conserved
\begin{equation}    \partial ^{\mu_1}E_{\mu_{1}\mu_{2}\ldots\mu_{n}\,|\,
\nu }=0 , \qquad    \partial ^\nu E_{\mu_{1}\mu_{2}\ldots \mu_{n-1 } \mu_{n} \, |\,
\nu }=0
\label{econs}
\end{equation}
as a consequence of (\ref{sbo}),(\ref{sft}).
This can be seen by rewriting $E$ in a form similar to (\ref{eindel}):
\begin{equation}
 E^{\mu_{1}\mu_{2}\ldots\mu_{n} |} {}_{\nu}= \,
 \delta ^{\mu_{1}\mu_{2}\ldots\mu_{n} \alpha\beta } _ {\rho_{1}\rho_{2}\ldots \rho_{n+1} \nu} \,\,S^{\rho_{1}\rho_{2}\ldots \rho_{n+1} \, |\,
 }{}_{\alpha\beta }
 \label{seindel}
\end{equation}
so that (\ref{econs}) immediately follows from (\ref{sft}). Note that $E_{\mu_{1}\mu_{2}\ldots\mu_{n}\,|\,
\nu }$ is an $[n,1]$ tensor satisfying
\begin{equation}
E_{[\mu_{1}\mu_{2}\ldots\mu_{n}\,|\,
\nu] }=0
\label{esymmis}
\end{equation}

Then the dual graviton can be coupled to an $[n,1]$ tensor source
$U_{\mu_{1}\mu_{2}\ldots\mu_{n}\, |\,
\nu}$ 
satisfying
\begin{equation}
U_{[\mu_{1}\mu_{2}\ldots\mu_{n}\, |\,
\nu]}=0
\label{antisym}
\end{equation}
through
\begin{equation}E_{\mu_{1}\mu_{2}\ldots\mu_{n}  \, |\,
 \nu}=U_{\mu_{1}\mu_{2}\ldots\mu_{n}   \, |\,
 \nu }
\label {duein}
\end{equation}
provided the source is conserved
\begin{equation}    \partial ^{\mu_1}U_{\mu_{1}\mu_{2}\ldots\mu_{n} \, |\,
\nu }=0 , \qquad    \partial ^\nu U_{\mu_{1}\mu_{2}\ldots\mu_{n} \, |\,
\nu }=0
\label {ucons}
\end{equation}
The action is \cite{deMedeiros:2002qpr}
\begin{equation}
S=\int d^dx \left(
 \frac 1 2 D^{\mu_{1}\mu_{2}\ldots\mu_{n} \, |\,
\nu }
 E_{\mu_{1}\mu_{2}\ldots\mu_{n} \, |\,
\nu }
-D^{\mu_{1}\mu_{2}\ldots\mu_{n} \, |\,
\nu }
U_{\mu_{1}\mu_{2}\ldots\mu_{n} \, |\,
\nu }
\right)
\end{equation}
and is  invariant under the gauge transformations (\ref{gaugvar}) as a result of (\ref{econs}) and (\ref{ucons}).
It will be useful to introduce the 
 trace $U'$ of the current $U$  which is an $(n-1)$-form
\begin{equation}U'_{\mu_{1}\mu_{2}\ldots\mu_{n-1}}= U_{\mu_{1}\mu_{2}\ldots\mu_{n} \, |\, \nu} \, \eta ^{\nu \mu_n}
\label {upr}
\end{equation}
Then (\ref {duein}) implies
\begin{equation}S''_{\mu_{1}\mu_{2}\ldots\mu_{n-1}}=-
U'_{\mu_{1}\mu_{2}\ldots\mu_{n-1}}
\end{equation}

Using the duality relation (\ref{sis}), the current $U$ gives a source for $R_{[\mu\nu \, |\, \sigma]\tau}$ so that   the graviton Bianchi identity 
(\ref{rbo}) no longer holds:
\begin{equation}  R_{[\mu\nu \, |\, \sigma]\tau} -R_{[\mu\nu \, |\, \sigma\tau ]}=\frac 1 {n!}
\epsilon_{\mu\nu \sigma}{}^
{\rho_1\rho_2 \dots \rho_n}  U_{\rho_1\rho_2 \dots \rho_n \, |\, \tau}
\label {fesoura}
\end{equation}
or equivalently
\begin{equation}
   R_{[\mu\nu \, |\, \sigma]\tau} =\frac 1 {n!}
\epsilon_{\mu\nu \sigma}{}^
{\rho_1\rho_2 \dots \rho_n} \bar U_{\rho_1\rho_2 \dots \rho_n \, |\, \tau}
\label{feqsa}
\end{equation}
where
\begin{equation}
\bar U_{\rho_1\rho_2 \dots \rho_n \, |\, \tau}= -(-1)^n
\frac 2 {3(n+2)}
\left(
U_{\rho_1\rho_2 \dots \rho_n \, |\, \tau}- \frac n 2 U'_{[\rho_1\rho_2 \dots \rho_{n-1}}\eta_ {\rho_n ]\tau}
\right)
\end{equation}
with
$U'$ given by (\ref{upr}).
Thus the current $U$ can be regarded as a magnetic source that leads to a violation of a Bianchi identity.
Similarly, $T_{\mu \nu} $ is, via the duality relation (\ref{sis}), a magnetic source for the dual gauge field $D$ giving a violation of the first Bianchi identity in 
(\ref{sbo}).

Generally, the case in which both sources $T,U$ are non-zero can be formulated in terms of a field strength ${\cal R} _{\mu\nu \rho\sigma}$ in the (reducible) $[2,0]\times [2,0]$ representation in which the trace ${\cal R} _{\mu\nu }$ is determined by $T$ and ${\cal R} _{[\mu\nu \rho]\sigma}$ is determined by $U$.
In regions in which $U=0$, ${\cal R} _{\mu\nu \rho\sigma}$ can be expressed in terms of a gauge field $h_{\mu \nu}$ and the theory can be formulated as conventional linearised gravity, while in regions in which $T=0$ it can be expressed in terms of a gauge field  $D_{\mu_{1}\mu_{2}\ldots\mu_{n}\, |\, \nu}$ giving the dual graviton theory.
In regions in which both sources are zero, both formulations are possible and are dual to each other. The $d=4$ theory with both sources has been discussed in \cite{Hull:2000rr,Bunster:2006rt,Barnich:2008ts}; the general case will be discussed elsewhere.

\section{ Conserved Charges}    \label{Conserved Charges}

\subsection{Conserved Charges for the Graviton}
 \label{Conserved Charges Grav}
 
 The graviton is invariant ($\delta h_{\mu \nu}=0$) under gauge transformations (\ref{hvar}) where the parameter $\xi$ is a  Killing vector of  Minkowski space $\xi_\mu=k_\mu$ satisfying
\begin{equation}
  \partial_{(\mu}k_{\nu)}=0
\label{killlin}
\end{equation}
Corresponding to the invariance under transformations generated by a given Killing vector $k$, there is a corresponding conserved charge $Q[k]$.
Indeed, 
\begin{equation}j_\mu [k]= T_{\mu\nu}k^\nu
\label{jtis}
\end{equation}
is a conserved current (using $  \partial^\mu T_{\mu\nu}=0$)
\begin{equation}  \partial^\mu j _\mu=0\end{equation}
and so can be integrated over a constant time slice $\Sigma$ to give a conserved charge
\begin{equation}Q[k]= \int _\Sigma j_\mu [k] \, d\Sigma ^\mu
\label{qkis}
\end{equation}
Using the field equation (\ref{eins}) 
 the current (\ref{jtis}) becomes
 \begin{equation}j_\mu [k]= G_{\mu\nu}k^\nu
\label{jtiss}
\end{equation}
so that the charge (\ref{qkis}) can be written as
\begin{equation}Q[k]= \int _\Sigma G_{\mu\nu}k^\nu \, d\Sigma ^\mu
\label{qkiss}
\end{equation}

\subsection{Conserved Charges for the Dual Graviton}

The  dual  graviton $D_{\mu_{1}\mu_{2}\ldots\mu_{n} \, |\, \nu}$ is invariant ($\delta D=0$) under gauge transformations (\ref{gaugvar}) with parameter $\alpha$
given by a generalised Killing tensor $\kappa  _{\mu_1\ldots \mu_{n-1}  \, |\, \rho}$ which is in the  $[n-1,1]$ representation and satisfies
\begin{equation}
  \partial _{[\mu} \kappa   _{\nu\ldots \sigma]\,|\rho} =0
\label {kkill}
\end{equation}
or transformations (\ref{gaugvar}) with parameter $\beta$
given by a  tensor $\lambda _{\mu_1\ldots \mu_n}$  in the $[n]$ representation satisfying
\begin{equation}
  \partial _ \rho \lambda  _{\mu \nu \ldots \sigma}-  \partial _ {[\rho} \lambda  _{\mu \nu
\ldots
\sigma]}
=0
\label {lkill}
\end{equation}
so that $\lambda _{\mu_1\ldots \mu_n}$
is a Killing-Yano tensor.

For each Killing tensor $\kappa  $ there is a  current
\begin{equation} j_\rho[\kappa  ]= \frac 1 {(n-1)!}  \kappa   ^{\mu_1\ldots \mu_{n-1}  \, |\, \nu}
U_{\rho \mu_{1}\mu_{2}\ldots\mu_{n-1}   \, |\, \nu }
\label {kkis}
\end{equation}
which is conserved ($  \partial _\mu j^\mu=0$) as a consequence of (\ref{ucons}),(\ref{kkill}), and for each Killing-Yano tensor $\lambda $ there is a  current
\begin{equation} 
j_\nu [\lambda ]= 
\frac 1 {n!} \lambda  ^ {\mu_1\ldots \mu_n} U_{ \mu_{1}\mu_{2}\ldots\mu_{n}   \, |\, \nu }
\label{llis}
\end{equation}
which is conserved as a consequence of (\ref{antisym}),(\ref{ucons}),(\ref{lkill}).
Integrating these over a constant time slice gives conserved charges $Q[\kappa  ],Q[\lambda ]$:
\begin{equation}Q[\kappa  ]=
  \int_\Sigma \, 
  j_\rho[\kappa  ]
 \,
d\Sigma ^\rho
\label{Qkkis}
\end{equation}
\begin{equation}
Q[\lambda ]=
 \int_\Sigma \, 
  j_\rho[\lambda  ]
 \,
d\Sigma ^\rho
\label{Qllis}
\end{equation}

The 1-form current 
$ j_\rho[\kappa  ]$
can be viewed as the Noether current for the 0-form symmetry with parameter 
$\varepsilon$
given by (\ref{gaugvar})   with
$\alpha _{\nu\ldots \sigma \,|\,\rho} =\varepsilon\kappa _{\nu\ldots \sigma \,|\,\rho} $
while the 1-form current 
$ j_\rho[\lambda  ]$
can be viewed as the Noether current for the 0-form symmetry  
with parameter $\epsilon $
given by (\ref{gaugvar}) with $\beta _{\mu \nu \ldots \sigma}
 =\epsilon \lambda _{\mu \nu \ldots \sigma}$.
 Then $Q[\kappa  ],Q[\lambda ]$ are the corresponding Noether charges.

Using the field equation (\ref{duein}), the currents can be written as
\begin{equation}
j_\rho[\kappa  ]= \frac 1 {(n-1)!}  \kappa   ^{\mu_1\ldots \mu_{n-1}  \, |\, \nu}
E_{\rho \mu_{1}\mu_{2}\ldots\mu_{n-1}   \, |\, \nu }
\label{kkise}
\end{equation}
and \begin{equation}
j^\nu [\lambda ]= 
\frac 1 {n!} \lambda  _ {\mu_1\ldots \mu_n} E^{ \mu_{1}\mu_{2}\ldots\mu_{n}   \, |\, \nu }
\label{llise}
\end{equation}
The relation (\ref{seindel}) gives
\begin{equation}
j^\rho[\kappa  ]= \frac 1 {(n-1)!}   \,    \delta ^{\rho \mu_{1}\mu_{2}\ldots\mu_{n-1} \alpha\beta } _ {\rho_{1}\rho_{2}\ldots \rho_{n+1} \nu} \,\, 
\kappa   _{\mu_1\ldots \mu_{n-1} }{}^ \nu
 \,S^{\rho_{1}\rho_{2}\ldots \rho_{n+1} }{}_{\alpha\beta }
 \label{curist}
\end{equation}
The dual of the Killing tensor 
\begin{equation}\tilde \kappa    ^{ \alpha \beta  \gamma \delta\, |\,  \nu}
=
 \frac 1 {(n-1)!}  
\kappa   _{\mu_1 \mu_2\ldots \mu_{n-1}\,|}{}^{ \nu}
\,
\epsilon ^{\mu_{1}\mu_{2}\ldots\mu_{n-1} \alpha \beta \gamma \delta} 
\label{dukapis}
 \end{equation}
satisfies
\begin{equation}
\tilde \kappa    ^{ \alpha \beta  \gamma \delta\, |\,  \nu}
=\tilde \kappa    ^{[ \alpha \beta  \gamma \delta ]\, |\,  \nu}
,\qquad 
  \partial _\alpha \tilde \kappa    ^{ \alpha \beta  \gamma \delta\, |\,  \nu}
=0, \qquad  \tilde \kappa    ^{ \alpha \beta  \gamma \delta }{}_\alpha=0
\label{kapids}
 \end{equation}
Note that $\tilde \kappa    ^{ [\alpha \beta  \gamma \delta\, |\,  \nu ]}
$ need not vanish. 
The current (\ref{curist}) can be rewritten (using (\ref{epep}) with $r=1$) as
\begin{equation} j^\alpha[\kappa  ]=- (-1)^n
 \frac 1 {d-1} \,   
\tilde \kappa ^{\alpha  \nu \rho \sigma
 | \mu}
 \,
\tilde S_{\mu[ \nu \rho \sigma ] }
 %S^{\rho_{1}\rho_{2}\ldots \rho_{n+1} }{}_{\alpha\beta }
 \label{curistaaa}
\end{equation}
where $\tilde S(D) $
is the dual of $S(D)$
\begin{equation}S_{\mu_{1}\mu_{2}\ldots\mu_{n+1}\;
\nu\rho }
\equiv \frac 1 2
\epsilon _{\mu_{1}\mu_{2}\ldots\mu_{n+1}\alpha\beta}
\tilde S^{\alpha\beta }{}_{\nu\rho}
\label{sisty}
\end{equation}
Gravitational duality amounts to identifying $\tilde S(D) $ with $R(h)$, or equivalently $  S(D) $ with $*R(h)$.

Next consider the current (\ref{llise})
which can be rewritten using (\ref{seindel}) as
\begin{equation} j_\nu [\lambda ]=  \frac 1 {n!} \lambda  _ {\mu_1\ldots \mu_n} 
 \delta ^{\mu_{1}\mu_{2}\ldots\mu_{n} \alpha\beta } _ {\rho_{1}\rho_{2}\ldots \rho_{n+1} \nu} \,\,S^{\rho_{1}\rho_{2}\ldots \rho_{n+1} \, |\,
 }{}_{\alpha\beta }
\label{lliser}
\end{equation}
Introducing the dual of the  Killing-Yano $n$-form $\lambda$, 
\begin{equation} 
\tilde  \lambda ^ {\alpha\beta\gamma}= \frac 1 {n!} \epsilon ^
{ \alpha\beta 
\gamma
{\mu_{1}\mu_{2}\ldots\mu_{n} }}
\lambda  _ {\mu_1\ldots \mu_n} 
\end{equation}
the current can be written as
\begin{equation} j_\nu [\lambda ]=  (-)^n \frac 1{d-1} 
\tilde  \lambda ^ {\alpha\beta\gamma}
\,\, 
\tilde S_{\nu [ \alpha\beta \gamma ] }
 \label{llisers}
\end{equation}

%%%%
\subsection{Killing Vectors }

The general Killing vector in Minkowski space is of the form
\begin{equation}
k_\mu=V_\mu +\Lambda_{\mu\nu}x^\nu
\label{KVec}
\end{equation}
for  a constant vector  $V_\mu$ and constant 2-form, $\Lambda_{\mu\nu}=-\Lambda_{\nu\mu}$.
Then
\begin{equation}
Q[k]= V_\mu P^\mu +\frac 1 2 \Lambda_{\mu\nu} J^{\mu\nu}
\label{momang}
\end{equation}
where $P^\mu$ is the momentum and $J^{\mu\nu}$ is the angular momentum. In particular,   $P^0$ is the mass.
As the symmetries (\ref{hvar}) are abelian, the corresponding algebra of the charges $P^m,J^{mn}$ is abelian in the linearised theory.

\subsection{Killing Tensors for the Dual Graviton}

A Killing-Yano tensor     $\lambda _{\mu_1\ldots \mu_n}$  in the $[n]$ representation satisfies
\begin{equation}
  \partial _ \rho \lambda  _{\mu \nu \ldots \sigma}-  \partial _ {[\rho} \lambda  _{\mu \nu
\ldots
\sigma]}
=0
\label {lkilla}
\end{equation}
The general Killing-Yano tensor satisfying (\ref{lkilla}) in Minkowski space 
is 
\begin{equation}
\lambda  _{\mu_1\ldots \mu_n}=
a_{\mu_1\ldots \mu_n}+ b_{\mu_1\ldots \mu_n \nu}x^\nu
\label {lkilt}
\end{equation}
where $a_{\mu_1\ldots \mu_n}$,  $b_{\mu_1\ldots \mu_{n +1}}$ are constant totally antisymmetric tensors, and so consist of an $n$-form and an 
$(n+1)$-form.
This then gives an
$n$-form charge $\hat P _ {\mu_1\ldots \mu_n}$ and an $n+1$-form charge  $\hat J _{\mu_1\ldots \mu_{n +1} }$
given by 
\begin{equation}
Q[\lambda]=
\frac 1 {n!}
 a_{\mu_1\ldots \mu_n}
 \hat P ^ {\mu_1\ldots \mu_n}
 +\frac 1 {(n+1)!}
 b_{\mu_1\ldots \mu_{n +1}}
  \hat J ^{\mu_1\ldots \mu_{n +1}}
  \label {maghat}
\end{equation}

Consider now the generalised Killing tensor $\kappa  _{\mu_1\ldots \mu_{n-1}  \, |\, \rho}$  in the  $[n-1,1]$ representation   satisfying
\begin{equation}
  \partial _{[\mu} \kappa   _{\nu\ldots \sigma]\rho} =0
\label {kkilla}
\end{equation}
A particular class of  \lq closed'   $[n-1,1]$ Killing tensors $\kappa  $ satisfying (\ref{kkilla})  are  those that are  \lq exact', i.e.\  of the form
\begin{equation}
\kappa   _{\mu_1\ldots \mu_{n-1}  \, |\, \rho}= 
  \partial _{[\mu_1} \sigma _{\mu_2\ldots \mu_{n-1} ]  \, |\, \rho}
  \label {kaptriv}
\end{equation}
for $n>2$.
Such \lq exact'  tensors automatically satisfy (\ref{kkilla}) for any 
$[n-2,1]$ tensor $\sigma _{\mu_1\ldots \mu_{n-2}  \, |\, \rho}$.
As $\sigma _{\mu_1\ldots \mu_{n-2}  \, |\, \rho}$ is an
$[n-2,1]$ tensor it satisfies 
$\sigma _{[\mu_1\ldots \mu_{n-2}  \, |\, \rho]}=0$ and so
(\ref{kaptriv}) satisfies
$\kappa   _{[\mu_1\ldots \mu_{n-1}  \, |\, \rho]}=0$ and as a result is an 
$[n-1,1]$ tensor.
For such Killing tensors, the charge 
(\ref{Qkkis}) becomes a boundary term
\begin{equation}Q[\kappa  ]=
\frac 1 {(n-1)!}  
 \int _S\, 
 \sigma ^{\mu_1\ldots \mu_{n-2}  \, |\, \nu}
U_{\rho \sigma \mu_{1}\mu_{2}\ldots\mu_{n-2}   \, |\, \nu }
\,
d\Sigma ^{\rho \sigma}
\label{Qkkisb}
\end{equation}
which will vanish if $  U $ vanishes on $S$. Then, when $  U $ vanishes on $S$,  the charge $Q[\kappa  ]$ is determined by an equivalence class of Killing tensors $K$ satisfying (\ref{kkill}) modulo exact pieces
\begin{equation}
\kappa   _{\mu_1\ldots \mu_{n-1}  \, |\, \rho}\sim \kappa   _{\mu_1\ldots \mu_{n-1}  \, |\, \rho}+
  \partial _{[\mu_1} \sigma _{\mu_2\ldots \mu_{n-1} ]  \, |\, \rho}
\end{equation}
so that the charge is defined by a generalised cohomology class of bi-forms.
This reflects the reducibility of the gauge symmetry. 

For $n=2$,  the requirement that $\kappa_{\mu |\nu}$ is a $[1,1]$ tensor means that it is a symmetric tensor, $\kappa_{\mu |\nu}=\kappa_{(\mu |\nu)}$.
In this case, (\ref{kaptriv}) is not automatically a $[1,1]$ tensor as it is not necessarily symmetric.
For $n=2$, the trivial Killing tensors are those of the form
\begin{equation}
\kappa_{\mu |\nu}=\partial _\mu
 \partial _\nu \psi
\label {tyutr}
\end{equation}
for some scalar $\psi$. This is  symmetric and automatically satisfies  $\partial _{[\rho}\kappa_{\mu] |\nu}=0$. Such trivial Killing tensors again give trivial charges.

For any constant vector $V^\mu$, $\kappa   _{\mu_1\ldots \mu_{n-1}  \, |\, \rho} V^\rho$ is  a closed $n-1$ form $i_V\kappa  $ and the non-trivial charges arise when this form is not exact. A tensor  $\kappa   _{\mu_1\ldots \mu_{n-1}  \, |\, \rho} = \beta _{\mu_1\ldots \mu_{n-1} } V_\rho$ 
will be a Killing tensor if the $n-1$ form $\beta _{\mu_1\ldots \mu_{n-1} }$ is closed and will be  in the  $[n-1,1]$ representation if
$\beta _{[\mu_1\ldots \mu_{n-1} } V_{\rho]}=0$, i.e.\  $V\wedge \beta=0$, which is satisfied if $\beta=V\wedge \alpha$ for some closed $n-2$ form $\alpha$.
This then leads to Killing tensors of the form
\begin{equation}
\kappa   _{\mu_1\ldots \mu_{n-1}  \, |\, \rho}= \alpha
  _{[\mu_1\ldots \mu_{n-2} }V_ {\mu_{n-1}]}V_ \rho
  \label{vkap}
\end{equation}
for some closed $n-2$ form $\alpha$ and constant vector $V$.
The charge corresponding to the tensor (\ref{vkap}) will be denoted
\begin{equation}
\label{asdert}
Q[\kappa]=Q[V,\alpha]\end{equation}

For constant $\kappa$, it will be useful to write
\begin{equation}
Q[\kappa] 
=\frac 1 {(n-1)!}\tilde P ^{\mu_1\ldots  \mu_{n-1}  |\rho}
\kappa_{\mu_1\ldots    \mu_{n-1}  |\rho}
 \label {pduaax}
\end{equation}
For $\kappa$ of the form (\ref{vkap})
with constant $\alpha$, this is
\begin{equation}
Q[V,\alpha]
=\frac 1 {(n-1)!}\tilde P ^{\mu_1\ldots \mu_{n-2}  \mu_{n-1}  |\rho}
 \alpha _{[\mu_1\ldots \mu_{n-2} }V_ {\mu_{n-1}]}V_ \rho
 \label {pduaa}
\end{equation}

Another Killing tensor is
\begin{equation}
\kappa   _{\mu_1\ldots \mu_{n-1}  \, |\, \rho}= 
\rho _{[\mu_1\ldots \mu_{n-2}} \eta _{\mu_{n-1} ]\rho}
\label{rhois}
 \end{equation}
where $\rho _{\mu_1\ldots \mu_{n-2}}$ is a closed $n-2$ form. and there is no dependence on any vector.
(This can be thought of as  linear combination of  tensors of the form (\ref{vkap})  for different choices of $V$.)
In this case, the dual is totally antisymmetric
\begin{equation}
\tilde \kappa    ^{ \alpha \beta  \gamma \delta\, |\,  \nu}
=\tilde \kappa    ^{[ \alpha \beta  \gamma \delta \, |\,  \nu]}\propto (*\rho) ^{ \alpha \beta  \gamma \delta    \nu}
\label{kapidsa}
 \end{equation}
so that the current (\ref{kkise}) becomes
\begin{equation}
j_\nu[\kappa  ]
\propto \rho   ^{\mu_1\ldots \mu_{n-2}  }
S''_{ \nu \mu_{1}\mu_{2}\ldots\mu_{n-2}   }
\propto
 (*\rho) ^{\nu \alpha  \beta \gamma \delta}
 \,
\tilde S_{[\alpha  \beta \gamma \delta ] }
\end{equation}

Further solutions are obtained by replacing $ \eta _{\mu  \rho}$ in (\ref{rhois}) with any constant symmetric tensor $ S _{\mu  \rho}$
or more generally by taking
\begin{equation}
\kappa   _{\mu_1\ldots \mu_{n-1}  \, |\, \rho}= 
\tau _{[\mu_1\ldots \mu_{r}}C_{ \mu_{r+1}
\dots \mu_{n-1} ]\rho}
\label{tauis}
 \end{equation}
 where $\tau _{\mu_1\ldots \mu_{r}}$ is a closed $r$-form and $C_{ \mu_{1}
\dots \mu_{n-r-1} |\rho}$ is a constant $[n-r-1,1]$ tensor so that $C_{ \mu_{1}
\dots \mu_{n-r-1} |\rho}=C_{[ \mu_{1}
\dots \mu_{n-r-1}] |\rho}$
and $C_{[ \mu_{1}
\dots \mu_{n-r-1} |\rho ]}=0$.

\section{Surface Integral Charges}\label{Surface}

Each of the currents considered in section \ref{Conserved Charges} can be written as a surface integral by first using the field equations to express them in terms of gauge fields and then finding a secondary 2-form current $J_{\mu \nu}$
such that
$j_\mu =   \partial ^\nu J_{\mu \nu}$
so the corresponding charge can be written as an integral over the 
$(d-2)$ dimensional boundary $S=  \partial \Sigma$
$$Q=\int _\Sigma j_\mu \, d  \Sigma^{\mu }= \frac 1 2  \int _S \, J _{\mu \nu} \, d \Sigma^{\mu \nu}$$
The existence of secondary 2-form currents and their explicit form follows from the arguments of \cite{Barnich:2001jy,Lee:1990nz,Wald:1999wa},  but for the linearised theory it is straightforward to find them directly, following \cite{Abbott:1981ff}.
Note that $j_\mu =   \partial ^\nu J_{\mu \nu}$
does not determine $J$ uniquely. Given any solution $J$, another is given by $${J'}_{\mu \rho}=J _{\mu \rho}+ \partial^ \sigma \Lambda _{\mu \rho \sigma}$$
where $\Lambda _{\mu \rho \sigma}=\Lambda _{[\mu \rho \sigma]}$ is an arbitrary 3-form. The corresponding charge is independent of $\Lambda _{\mu \rho \sigma}$ as $Q=\int _S*J=\int_S *J'$.

\subsection{The ADM charges $Q[k]$}
\label{gravchar}

Consider first the  charges
$Q[k]$ given by (\ref{qkis}).
Using the field equation (\ref{eins})  
 the current (\ref{jtis}) becomes
 \begin{equation}j_\mu [k]= G_{\mu\nu}k^\nu
\label{jtiss}
\end{equation}
 If there are no magnetic sources, i.e\ if $U=0$, then  (\ref{ris}) holds and, using the identity (\ref{eindel}), the Einstein tensor  $G_{\mu \nu} $ can   be written as
\begin{equation} G^{\mu \nu}=   \partial _\rho   \partial _\sigma K^{\mu \rho  \, |\,\nu \sigma}
\end{equation}
where
\begin{equation} K^{\mu \rho  \, |\,\nu \sigma}=-%6
3 
\eta ^{\mu \rho \alpha | \nu \sigma \beta} h_{\alpha \beta}
\label{kist}
\end{equation}
with
\begin{equation} 
\eta ^{\mu \rho \alpha | \nu \sigma \beta} \equiv  \eta^{  \mu \tau} \eta^{ \rho \lambda} \eta^{ \alpha \gamma}
 \delta _{\tau \lambda \gamma}^{\nu \sigma \beta} 
\end{equation}
Note that $K^{\mu \rho  \, |\,\nu \sigma}$ has the algebraic properties of the Riemann tensor and the identity $\partial _\mu G^{\mu \nu}=0$ clearly follows. This can now be used to write the current (\ref{jtiss}) as a total derivative
 \begin{equation}j^\mu [k]= (  \partial _\rho   \partial _\sigma K^{\mu \rho  \, |\,\nu \sigma} )k_\nu=
   \partial _\rho [(  \partial _\sigma K^{\mu \rho  \, |\,\nu \sigma} )k_\nu - 
  K^{\mu \sigma  \, |\, \nu  \rho}   \partial _\sigma k_\nu]
\label{surk}
\end{equation}
where the fact that the Killing vectors $k$ in Minkowski space satisfy
\begin{equation} 
   \partial _\rho   \partial _\sigma  k_\nu=0
 \label{ddkis}
 \end{equation}
has been used.
The current  is then of the desired form $j_\mu [k]=   \partial ^\nu J _{\mu \nu}[k]$
where
\begin{equation}
J ^{\mu \rho}[k]=(  \partial _\sigma K^{\mu \rho  \, |\, \nu \sigma} )k_\nu - 
  K^{\mu \sigma   \, |\,\nu  \rho}   \partial _\sigma k_\nu% +  \partial_ \sigma \Lambda ^{\mu \rho \sigma}
\end{equation}
The corresponding charge is  
\begin{equation} 
Q[k]=
\frac 1 2  \int _S\,   \, d \Sigma_{\mu \rho} \left[ (  \partial _\sigma K^{\mu \rho  \, |\,\nu \sigma} )k_\nu - 
  K^{\mu \sigma  \, |\,\nu  \rho}   \partial _\sigma k_\nu \right]
  \label{ADM}
\end{equation}
%and the integral is independent of $\Lambda ^{\mu \rho \sigma}$. 
This gives the linearised form of the standard ADM expressions for momentum and angular momentum.

Using (\ref{conne}),(\ref{kist}) 
\begin{equation}
 \partial _\sigma K^{\mu \rho  \, |\,\nu \sigma}=
 -3 \eta ^{\mu \rho \alpha | \nu \sigma \beta} \Gamma_{\sigma\beta \, \alpha }
\end{equation}
so that $Q[k]$ can be written as
\begin{equation} 
Q[k]=
-\frac 3 2 \int _S\,   \, d \Sigma_{\mu \rho} \left[ 
\eta ^{\mu \rho \alpha | \nu \sigma \beta}  \Gamma_{ \sigma\beta\, \alpha}
  k_\nu - 
  \eta ^{\mu \sigma \alpha | \nu  \rho \beta} h_{\alpha \beta}
     \partial _\sigma k_\nu \right]
  \label{ADMa}
\end{equation}
For a translation Killing vector $k^\mu$ is a constant vector so that the second term in (\ref{ADMa})
vanishes to leave
\begin{equation}
Q[k]  =- %\frac 3 2
\frac 3 2
\int _S\delta _{\mu\nu\rho}^{\alpha\beta\gamma} \, k ^{\mu}\, \Gamma^  {\nu\rho} {}_  {\gamma}
 \, d\Sigma _{\alpha\beta}
 \label{nester}
\end{equation}
As $k^\mu$ is a constant vector this gives the momentum vector $P^\mu$ via  $Q[k] =k^\mu P_\mu$.
This is the linearised form of the expression for the momentum   given in \cite{Nester:1981bjx}.
For the killing vector in the time direction $k^\mu=\delta ^\mu_0$ the charge gives the mass $M$ $Q[k]=M$.
With coordinates $x^\mu=(x^0,x^I)$ where $I=1,\dots d-1$ labels the spatial directions,
the expression (\ref{ADMa}) reduces to
\begin{equation}
M=-\frac 1 2 \int _S (h_{IJ,J}-h_{JJ,I})
d\Sigma ^{0I}
\label{sadm}
\end{equation}

\subsection{The dual charges $Q[\kappa] $}

Consider next the charge (\ref{Qkkis}).
The current (\ref{kkis}) becomes 
\begin{equation}
j_\rho[\kappa  ]= \frac 1 {(n-1)!}  \kappa   ^{\mu_1\ldots \mu_{n-1}  \, |\, \nu}
E_{\rho \mu_{1}\mu_{2}\ldots\mu_{n-1}   \, |\, \nu }
\label{kkise}
\end{equation}
using the field equation (\ref{duein}).
If there are no electric sources, i.e.\ if $T_{\mu\nu}=0$, then the dual field strength $S$ is given by (\ref{dufistr})
 and, using the identity (\ref{seindel}), the dual Einstein tensor $E$
 can   be written as
 \begin{equation}
 E^{\mu_{1}\mu_{2}\ldots\mu_{n}  \, |\,\nu}=
    \partial _\rho   \partial _\sigma  L ^{\mu_{1}\mu_{2}\ldots\mu_{n} \rho} {}^{| \nu \sigma}
  \label{seindell}
\end{equation}
 where
\begin{equation}
    L ^{\mu_{1}\mu_{2}\ldots\mu_{n} \rho} {}^{| \nu \sigma}=
     \,
 \eta^{\lambda \mu_{1}\mu_{2}\ldots\mu_{n}  \rho  } {}^ {|   \tau_{1}\tau_{2}\ldots \tau_{n} \sigma\nu} \,\,
 D_{\tau_{1}\ldots \tau_{n}  \, |\,\lambda }    \label{seindellaa}
\end{equation}
 Using (\ref{kkilla}),
 the current (\ref{kkise})  becomes
\begin{eqnarray}
j^\tau[\kappa  ]
%&=& \frac 1 {(n-1)!}  \kappa   _{\mu_1\ldots \mu_{n-1}  \, |\, \nu} E^{\tau \mu_{1}\mu_{2}\ldots\mu_{n-1}   \, |\, \nu }  \\
&=&
\frac 1 {(n-1)!}  \kappa   _{\mu_1\ldots \mu_{n-1}  \, |\, \nu}
   \partial _\rho   \partial _\sigma  L^{\tau \mu_{1}\mu_{2}\ldots\mu_{n-1} \rho \, | \, \nu \sigma}\nonumber
 \\
&=&
\frac 1 {(n-1)!}   \partial _\rho \left[ \kappa   _{\mu_1\ldots \mu_{n-1}  \, |\, \nu}
   \partial _\sigma  L^{\tau \mu_{1}\mu_{2}\ldots\mu_{n-1} \rho \, | \, \nu \sigma}
% - L^{\tau \mu_{1}\mu_{2}\ldots\mu_{n-1} \sigma \, |\, \nu \rho}   \partial _\sigma  \kappa   _{\mu_1\ldots \mu_{n-1}  \, |\, \nu}
 \right]
 \nonumber
\end{eqnarray}
As a result, 
the corresponding 2-form current with 
$$j_\mu[\kappa] =   \partial ^\nu J_{\mu \nu}[\kappa]$$
 is
\begin{equation}
J^{\tau\rho}[\kappa]= \frac 1 {(n-1)!}    \kappa   _{\mu_1\ldots \mu_{n-1}  \, |\, \nu}
   \partial _\sigma  L^{\tau \mu_{1}\mu_{2}\ldots\mu_{n-1} \rho \, | \, \nu \sigma}
\end{equation}
with the   charge  given by the surface integral
\begin{equation}
 Q[\kappa]=
\frac 1 2  \int _S \, J _{\mu \nu} [\kappa]\, d \Sigma^{\mu \nu}
\end{equation}
The current $J^{\tau\rho}[\kappa]$  can be written in terms of $D$ as
\begin{equation}
J^{\tau\rho}[\kappa]= \frac 1 {(n-1)!}    \kappa   _{\mu_1\ldots \mu_{n-1}  \, |\, \nu}
    \eta^{\lambda \tau \mu_{1}\mu_{2}\ldots\mu_{n-1}  \rho  } {}^ {|   \alpha_{1}\alpha_{2}\ldots \alpha_{n} \sigma\nu} \,\,
 \partial _{[\sigma   }D_{\alpha_{1}\ldots \alpha_{n}]  \, |\,\lambda }\end{equation}
or written in terms of the connection (\ref{duconn}) using
\begin{equation}
     \partial _\sigma   L ^{\mu_{1}\mu_{2}\ldots\mu_{n} \rho} {}^{| \nu \sigma}=
 \eta^{\lambda \mu_{1}\mu_{2}\ldots\mu_{n}  \rho  } {}^ {|   \tau_{1}\tau_{2}\ldots \tau_{n} \sigma\nu} \,\,
   \tilde \Gamma _{\sigma\tau_{1}\ldots \tau_{n}  \, |\,\lambda } 
  %  \partial _\sigma  D_{\tau_{1}\ldots \tau_{n}  \, |\,\lambda }   
   \label{seindellaaa}
\end{equation}
  to give
  \begin{equation}
J^{\alpha\rho}[\kappa]=   \frac 1 {(n-1)!}    \kappa   _{\mu_1\ldots \mu_{n-1}  \, |\, \nu}
  \,
 \eta^{\lambda\alpha\mu_{1}\mu_{2}\ldots\mu_{n-1 }  \rho  } {}^ {|   \tau _{1}\tau_{2}\ldots \tau_{n} \sigma\nu} \,\,
  \tilde \Gamma _{\sigma \tau_{1}\tau_{2}\ldots\tau_{n} \, |  \lambda}
%   \partial _\sigma  L^{\tau \mu_{1}\mu_{2}\ldots\mu_{n-1} \rho \, | \, \nu \sigma}
\end{equation}
With the dual   $(*\tilde \Gamma ) _{\tau} {}^ {\mu\nu}(D)$ of the connection given by (\ref{ducond}),
the current $J[\kappa]$
 can  be rewritten in terms of the  dual Killing tensor (\ref{dukapis}) as
 \begin{equation}
J^{\alpha\rho}[\kappa]= (- )^{n}
\frac 1 {d-1}
\,
\tilde \kappa    ^{ \alpha \rho  \lambda     \gamma  \, |\,  \nu}
(*\tilde \Gamma ) _{\nu\gamma\, | \lambda } 
\label{Jtiisa}
\end{equation}
For a Killing tensor of the form (\ref{rhois}) this becomes
\begin{equation}
J^{\alpha\rho}[\kappa]
=
\frac 1 {(d-1 ) (d-4)}
\,
(* \rho )  ^{ \alpha \rho  \lambda     \gamma     \nu}
(*\tilde \Gamma ) _{[\nu\gamma\, | \lambda ]} 
\label{Jtiisar}
\end{equation}
and so vanishes if the \lq dual torsion' $(*\tilde \Gamma )  _{[\nu\gamma\, | \lambda ]} $ vanishes.

\subsection{The dual charges $Q[\lambda]$}

Consider next the charge  (\ref{Qllis}).
The current (\ref{llis}) becomes 
\begin{equation} 
j_\nu [\lambda ]= 
\frac 1 {n!} \lambda  _ {\mu_1\ldots \mu_n} E^{ \mu_{1}\mu_{2}\ldots\mu_{n}   \, |\, \nu }
\label{llisdfg}
\end{equation}
using the field equation (\ref{duein}).
If there are no electric sources, $T_{\mu\nu}=0$, then the dual field strength $S$ is given by (\ref{dufistr})
and $E$ is given by (\ref{seindell}) so that
\begin{eqnarray}
j^\nu [\lambda ]
&=& 
\frac 1 {n!}
  \partial _\rho  \left[
 \lambda  _ {\mu_1\ldots \mu_n}    \partial _\sigma  L ^{\mu_{1}\mu_{2}\ldots\mu_{n} \rho} {}^{| \nu \sigma}
 -
 L ^{\mu_{1}\mu_{2}\ldots\mu_{n} \sigma} {}^{| \nu \rho}   \partial _\sigma \lambda  _ {\mu_1\ldots \mu_n} 
 \right]
\end{eqnarray}
using the fact that the Killing-Yano tensors (\ref{lkilt}) satisfy  
\begin{equation}  \partial_\rho    \partial _\sigma  \lambda  _ {\mu_1\ldots \mu_n} 
=0
\label{ddlam}
\end{equation}
In Appendix A, it is shown that this can be rewritten as 
$$j_\mu[\lambda] =   \partial ^\nu J_{\mu \nu}[\lambda]$$
with
\begin{eqnarray}
J^{\nu\rho}[\lambda]&=&\frac 1 {n!}
\lambda  _ {\mu_1\ldots \mu_n}  \left[
     \eta^{\lambda \mu_{1}\mu_{2}\ldots\mu_{n}  \rho  } {}^ {|   \tau_{1}\tau_{2}\ldots \tau_{n} \sigma\nu} \,\,
  -
    \eta^{\lambda \mu_{1}\mu_{2}\ldots\mu_{n}     \nu } {}^ {|   \tau_{1}\tau_{2}\ldots \tau_{n} \sigma \rho} \,\,
  \right] \tilde \Gamma _{\sigma\tau_{1}\ldots \tau_{n}  \, |\,\lambda } 
  \nonumber
   \\
   &+& \frac n {n+1} \frac 1 {n!}
% L ^{\mu_{1}\mu_{2}\ldots\mu_{n} \sigma} {}^{| \nu \rho}  
 \partial _\sigma \lambda  _ {\mu_1\ldots \mu_n} 
  \eta^{\lambda \mu_{1}\mu_{2}\ldots\mu_{n} \sigma   } {}^ {| \nu    \rho \tau_{1}\tau_{2}\ldots \tau_{n}  } \,\,
 D_{\tau_{1}\ldots \tau_{n}  \, |\,\lambda }
\end{eqnarray}
For constant $\lambda$ the second line of this equation  vanishes
and $J$ can be rewritten as
\begin{equation}
J^{\nu\rho}[\lambda]=  (-1)^{n+1}\frac {1} {(n+2)}  \tilde \lambda^{  \alpha \beta\rho} (*\tilde \Gamma )_{\alpha }{}^\nu{}_{|\beta } -(\nu \leftrightarrow \rho)
\end{equation}
If the \lq torsion' $(*\tilde \Gamma )_{[\lambda |\alpha \nu] } =0$ this can be rewritten as
\begin{equation}
J^{\nu\rho}[\lambda]= (-1)^{n+1}
 \frac {1} {2(n+2)} 
 \tilde \lambda^{  \alpha \beta\rho} 
 (*\tilde \Gamma )_{\alpha\beta  }{}^{|\nu}
 -(\nu \leftrightarrow \rho)
\label{jlambloc}
\end{equation}
In Appendix A, it is shown that $j$ can also be written in terms 
 of $\hat \Gamma$ instead of $\tilde \Gamma$.

%%%%
\section{Magnetic Currents and Charges for Linearised Gravity}
\label{Currents and Charges Lin}

Secondary two-form currents $J[k],J[\kappa]$ and $J[\lambda ]$ have been constructed corresponding to the conserved 1-form primary currents $j[k],j[\kappa]$ and $j[\lambda]$.
As in each case $  \partial ^\mu J_{\mu \nu }=j_\nu$, the 2-form current $J$ is conserved in any region in which the corresponding $j=0$.
Then the charge defined
by
\begin{equation}
Q=\int_\Sigma *J
\end{equation}
over a $d-2$ dimensional surface $\Sigma$ is a topological operator in  the sense that $\Sigma$ 
 can be deformed to another surface $\Sigma'$ with $Q$ unchanged so long as no region with $j\ne 0$ is crossed.
In section \ref{Surface} expressions have been found for the currents $J[\kappa]$ and $J[\lambda]$ in terms of the dual graviton $D$. In this section, expressions for these currents given in terms of $h$ instead of $D$ will be found, and an expression for $J[k]$ in terms of $D$ instead of $h$ will be given.

In a region in which the source $U_{\mu_{1}\mu_{2}\ldots\mu_{n}\, |\,
\nu}=0$    the 1-form currents
$j[\kappa]$ and $j[\lambda]$ vanish so that  the
 2-form currents $J[\kappa]$ and $J[\lambda ]$ are  conserved identically. Moreover, in this region without magnetic sources the theory can be formulated  in terms of the graviton $h$.
In such a region, the current $J[\kappa]$  and the current $J[\lambda]$
for constant $\lambda$
can be written in terms of the graviton $h$, as will be discussed below, so they can be regarded as magnetic currents for this theory. The current $J[\lambda]$ for non-constant $\lambda$ depends explicitly on $D$ and so cannot be written locally  in terms of $h$. It   can be regarded as a non-local magnetic charge for gravity  as it is written in terms of the dual field $D$ which is non-locally related to $h$ by  (\ref{ris}). 
Conversely, in a region in which $T_{\mu\nu}=0$, the    theory can be formulated  in terms of the dual graviton $D$ and $j[k]$ vanishes so that $J[k]$ is identically conserved. The current $J[k]$ for constant $k^\mu$ (a translational Killing vector)  can be written in terms of the dual graviton $D$ and so is a magnetic current for the dual theory, whereas for non-constant $k$ (a rotation or boost Killing vector), its explicit dependence on $h$ means that it gives a non-local magnetic charge for the dual theory.

Consider  a region ${\cal U}$ in which $T_{\mu\nu}=0$ and $U_{\mu_{1}\mu_{2}\ldots\mu_{n}\, |\,
\nu}=0$. In ${\cal U}$, linearised gravity can be formulated in terms of the graviton $h$ with curvature $R(h)$ given by (\ref{ris})
or the dual graviton $D$ with field strength, $S(D)$ (see (\ref{dufistr})), and the field strengths are dual $S(D)=*R(h)$, (\ref{sis}). It is convenient to choose a gauge in which the connections $\Gamma(h)$ given by (\ref{conne}) and $\tilde \Gamma (D)$ given  by (\ref{duconn}) are dual, satisfying (\ref{ducon}).
The  current $J[\kappa]$ given by (\ref{Jtiisa}) can then be rewritten in terms of $\Gamma(h)$
as
 \begin{equation}
J^{\alpha\rho}[\kappa]= (- )^{d+1} \frac 1 {d-1}\,
\tilde \kappa    ^{ \alpha \rho  \lambda     \gamma  \, |\,  \nu}
\Gamma _{ \nu\gamma \, |\lambda }
\label{Jtiisbc}
\end{equation}
Then using
$\Gamma _{[ \nu\gamma\,|\lambda]}=0$ gives
 \begin{equation}
J^{\alpha\rho}[\kappa]= (- )^{d+1}  \frac 1 2  \frac 1 {d-1}\,  
\tilde \kappa    ^{ \mu\nu \rho \sigma        \, |\,  \lambda}
\Gamma _{  \rho \sigma\,|\lambda}
\label{Jtiisbcq}
\end{equation}
and using (\ref{conne}),(\ref{kapids}) this can be rewritten as
\begin{equation}
J_{\mu\nu}[\kappa]=\partial ^\rho Z_{\mu\nu\rho}[\kappa]
\end{equation}
where
\begin{equation}
 Z_{\mu\nu\rho}[\kappa]=(- )^{d+1}  \frac 1 2  \frac 1 {d-1}\, \tilde \kappa    _{ \mu\nu \rho \sigma        \, |  \tau}h^{\sigma\tau}
\end{equation}
This is manifestly conserved
  \begin{equation}
 \partial _\rho J^{\alpha\rho}[\kappa]= 0\label{Jtiisab}
\end{equation}
in the region ${\cal U}$.
The  charge is given by
the surface integral
\begin{equation}
 Q[\kappa]=
\frac 1 2  \int _S \, J _{\mu \nu} [\kappa]\, d \Sigma^{\mu \nu} =
\frac 1 2  \int _S \,  \partial ^\rho Z_{\mu\nu\rho}[\kappa] \, d \Sigma^{\mu \nu}
\label {Qkaph}
 \end{equation}
or as (\ref{Qkax1}) with the  $Z$ given by (\ref{Qkax2}).
The charge is then a total derivative and if $h_{\mu \nu}$ is globally defined it will be zero. It is then a topological charge reflecting the   non-triviality of  the $h_{\mu \nu}$ structure.

For the case in which $\kappa$ is of the form (\ref{rhois}),  using (\ref{ducon}) brings
(\ref{Jtiisar}) to 
\begin{equation}
J^{\alpha\rho}[\kappa]
\propto
 (- )^{n}
\frac 1 {d-1}
\,
\rho   ^{ \alpha \rho  \lambda     \gamma     \nu}
(  \Gamma ) _{[ \nu\gamma\, | \lambda]}
\label{Jtiisara}
\end{equation}
and
this vanishes as  the torsion (\ref{toris}) vanishes,  $\Gamma _{[ \nu\gamma\,|\lambda]}=0$.
Then %as the torsion vanishes 
the trace $\eta^{\nu\rho}\kappa  _{\mu_1\ldots \mu_{n-2} \nu \, |\, \rho}$ of $\kappa$ does not contribute to $ Q[\kappa]$ and
 $\kappa $ can be taken to be traceless, so that
$\tilde \kappa    ^{[ \alpha \rho  \lambda     \gamma  \, |\,  \nu]}=0$ and $\tilde \kappa $ is an irreducible $[4,1]$ tensor.

Similarly, the current   $J[\lambda]$
for constant $\lambda$ given by (\ref{jlambloc}) can be written in terms of $\Gamma(h)$ using the gauge choice (\ref{duconsdfs}) in which (\ref{ducon}) holds, giving 
\begin{equation}
J^{\nu\rho}[\lambda]= (-1)^n \frac 1 {2(n+2)} 
 \tilde \lambda^{  \alpha \beta\rho} 
   \Gamma _{\alpha\beta  }{}^\nu
 -(\nu \leftrightarrow \rho)
\label{jlamblocas}
\end{equation}
Defining
\begin{equation}
 Z_{\mu\nu\rho}[\lambda]= (-1)^n
\frac 1 {n+2} \tilde \lambda    _{[\sigma \mu\nu }h_{\rho]}{}^ \sigma   
\label{Zis}   
\end{equation}
gives
\begin{equation}
 \partial ^\rho Z_{\mu\nu\rho}[\lambda]= J_{\mu\nu}[\lambda]
%- \frac 1 { n+2} \tilde \lambda    _{\rho\sigma[\mu}\Gamma ^{\rho\sigma}{}_{|\nu]}
+(-1)^n\frac 1 {n+2} \tilde \lambda    _{ \mu\nu \sigma}\Gamma ^{\rho\sigma}{}_{ \rho}
\end{equation}
so that $J_{\mu\nu}[\lambda]\sim \partial ^\rho Z_{\mu\nu\rho}[\lambda]$
with the gauge   choice (\ref{duconsdfs}) $\Gamma ^{\rho\sigma}{}_{ \rho}=0$.

Relaxing the gauge condition (\ref{duconsdfs}), it will be convenient to {\it define}  $J_{\mu\nu}[\lambda]$ to be
\begin{equation}
J_{\mu\nu}[\lambda]=\partial ^\rho Z_{\mu\nu\rho}[\lambda]
\label {jjjiii}
\end{equation}
with $Z$ given by (\ref{Zis}).
This agrees with (\ref{jlamblocas}) in the gauge  (\ref{duconsdfs}) but is manifestly conserved
  \begin{equation}
 \partial _\rho J^{\alpha\rho}[\lambda]= 0\label{Jtiisabl}
\end{equation}
in the region ${\cal U}$ without the need to use the gauge condition (\ref{duconsdfs}).
As the conservation does not rely on the gauge choice 
this indeed defines a conserved current    $J^{\alpha\rho}[\lambda]$ in any gauge.
The  charge is 
\begin{equation}
 Q[\lambda]=
\frac 1 2  \int _S \, J _{\mu \nu} [\lambda]\, d \Sigma^{\mu \nu} =
\frac 1 2  \int _S \,  \partial ^\rho Z_{\mu\nu\rho} [\lambda]\, d \Sigma^{\mu \nu}
\label{Qlamh}
 \end{equation}
or as (\ref{Qkax3}) with the dual of $Z$ given by (\ref{Qkax4}).
This   is also a topological charge which vanishes if   $h_{\mu \nu}$ is globally defined.

As discussed above, the charge $Q[\lambda]$ for the dual graviton theory, in the case in which     $\lambda$ is non-constant and given by the term linear in $x$ in (\ref{lkilt}), cannot be immediately be rewritten for the graviton theory.
However, the current defined by (\ref{Zis}),(\ref{jjjiii})
is conserved for {\emph any} $\lambda$, and in particular for $\lambda$
given by the term linear in $x$ in (\ref{lkilt}). Then integrating this gives a new charge defined for the  non-constant $\lambda$. The relation between these two charges 
with $\lambda$ linear in $x$, one for the graviton theory and one for the dual graviton theory,
will be discussed in \cite{Hutt}.

Just as the charges $Q[\kappa]$ and  $Q[\lambda]$ can be interpreted as topological charges for the graviton theory of the field $h$, the charge momentum given by $Q[k]$ for constant $k$ can be regarded as topological charge for the dual graviton theory.
The charge (\ref{nester}) can be written in terms of $D$ using (\ref{ducon})  and (\ref{duconnanti}) to give
\begin{equation}
Q[k]=\int_S d Y
\end{equation}
where $Y$ is a $d-3$ form with components
\begin{equation}
Y_{\mu_{1}\mu_{2}\ldots\mu_{d-3}}= \frac 1 2 D_{\mu_{1}\mu_{2}\ldots\mu_{d-3}|\, \nu}k^\nu
\end{equation}

%%%%%%
%%%%%

%%%%%%%

\section{Four dimensions}	 \label{4DSec}

In four dimensions, the dual graviton is a symmetric tensor $D_{\mu\nu}=\tilde h_{\mu\nu}$ with gauge symmetry
\begin{equation} \delta \tilde h_{\mu\nu}=   \partial _{(\mu} \tilde \xi_{\nu)}
\label {hvart}
\end{equation}
The field strength $S_{\mu\nu\rho \sigma}[D]$ is the linearised curvature of $\tilde h_{\mu\nu}$
\begin{equation}
\tilde R_{\mu\nu\, \sigma\tau}= -2  \partial_{[\mu}\tilde h_{\nu][\sigma,\tau]}
\label {risti}
\end{equation}
The duality relation between the theory of $h$ and that of $\tilde h$ is the requirement that
the curvature  $S=\tilde R$ is the dual of $R$,
$\tilde R(\tilde h)=*R(h)$,
so that
\begin{equation}\tilde R_{\mu\sigma 
\nu\rho }
=\frac 1 2
\epsilon _{\mu\sigma \alpha\beta}
R^{\alpha\beta }{}_{\nu\rho}
\label{sisasa}
\end{equation}
The dual connection is
 \begin{equation}
 \tilde\Gamma _{  \mu\nu\, \tau}=   \partial _{[\mu }\tilde h_{\nu] \tau}
 \end{equation}
 and the duality relation (\ref{sisasa}) implies that one can choose a gauge in which this connection is dual to the connection (\ref{conne}):
  \begin{equation}
 \tilde\Gamma _{  \mu\nu\, \tau}=   \frac 1 2
\epsilon _{\mu\nu }{}^{ \alpha\beta}
\Gamma
_{  \alpha\beta\, \tau}
\label{ducon4}
 \end{equation}
 The source is a symmetric dual stress tensor $U_{\mu\nu}= \tilde T_{\mu\nu}$, with field equation given by 
 \begin{equation}
\tilde G_{\mu\nu}=\tilde T_{\mu\nu}
\label {4dein}
\end{equation}
where $\tilde G_{\mu\nu}$ is the linearised Einstein tensor for the dual graviton $\tilde h_{\mu\nu}$.

The   Killing tensors from gauge transformations preserving $\tilde h_{\mu\nu}$ are given by parameters $ \tilde \xi_{\nu}$ that are Killing vectors satisfying (\ref{killlin}).
The current 
\begin{equation}\tilde j_\mu [k]=\tilde T_{\mu\nu}k^\nu
\label{jtisti}
\end{equation}
is then   conserved, giving a conserved charge
\begin{equation}\tilde Q[k]= \int _\Sigma \tilde j_\mu [k] \, d\Sigma ^\mu
\label{qkistilde}
\end{equation}

The structure of the currents is of the same form as that described in sections \ref{Conserved Charges Grav} and \ref{gravchar}, but with $  h_{\mu\nu}$ replaced by the dual graviton $\tilde h_{\mu\nu}$.
Using the field equation (\ref{4dein}) for $\tilde h_{\mu\nu}$,
 the current  can be written as $\tilde j_\mu [k]=   \partial ^\nu \tilde J _{\mu \nu}[k]$
where
\begin{equation}
\tilde J ^{\mu \rho}[k]=(  \partial _\sigma \tilde  K^{\mu \rho  \, |\, \nu \sigma} )k_\nu - 
  \tilde K^{\mu \sigma   \, |\,\nu  \rho}   \partial _\sigma k_\nu 
\end{equation}
\begin{equation}
\tilde K^{\mu \rho  \, |\,\nu \sigma}=-6 \eta ^{\mu \rho \alpha | \nu \sigma \beta} \tilde h_{\alpha \beta}
\end{equation}
The corresponding charge is 
\begin{equation} 
\tilde Q[k]=
\frac 1 2  \int _S\,   \, d \Sigma_{\mu \rho} \left[ (  \partial _\sigma \tilde K^{\mu \rho  \, |\,\nu \sigma} )k_\nu - 
\tilde   K^{\mu \sigma  \, |\,\nu  \rho}   \partial _\sigma k_\nu \right]
  \label{ADMp}
\end{equation}

For constant Killing vectors $k$
the charge can be written in the form  (\ref{nester}), but with the connection $\Gamma$ replaced by the dual connection $\tilde \Gamma$
\begin{equation}
\tilde Q[k] =- \frac 3 2
\int _S\delta _{\mu\nu\rho}^{\alpha\beta\gamma} \, k ^{ \mu}\, \tilde\Gamma^  {\nu\rho} {} _  {\gamma}
 \, d\Sigma _{\alpha\beta}
 \label{nesterdu}
\end{equation}
This can then be rewritten in terms of $\Gamma(h)$  using (\ref{epep}),(\ref{ducon4}) as
\begin{equation}
\tilde Q[k] =- 
\int _S
\epsilon ^{\alpha\beta\gamma\delta } k^\mu \,  \Gamma_   {\mu\delta\, \gamma} 
 \, d\Sigma _{\alpha\beta}
 \label{nesterdudu}
\end{equation}
Using $\Gamma_  {[\gamma |\mu\delta]} =0$, this becomes
\begin{equation}
\tilde Q[k] =- \frac 1 2
\int _S
\epsilon ^{\alpha\beta\gamma\delta } k^\mu \,  \Gamma_  {\gamma\delta \, \mu} 
 \, d\Sigma _{\alpha\beta}
 =-\int _S
 k^\mu \,  (*\Gamma)_ {\alpha\beta\, \mu} 
 \, d\Sigma ^{\alpha\beta}
 \label{nesterdudu}
\end{equation}
The charges $\tilde P^\mu$ defined for constant $k^\mu$ by
$\tilde Q[k]=k_\mu \tilde P^\mu$ is the linearised version of the dual momentum or NUT 4-momentum introduced for general relativity in \cite{Ramaswamy,Ashtekar}. The charge can be written as
\begin{equation}
 \tilde Q[k] =
\frac 1 2  \int _S \,  \partial ^\rho Z_{\mu\nu\rho} (k)\, d \Sigma^{\mu \nu}
 \end{equation}
where
\begin{equation}
  Z_{\mu\nu\rho}(k)=-\epsilon_{\mu\nu\rho \sigma}k^\tau h_\tau {}^\sigma
\end{equation}
and so is a topological charge that is zero if $h_{\mu\nu}$ is a  globally defined tensor.
It can be written as
\begin{equation}
 \tilde Q[k] =
\frac 1 2  \int _S \,  d (*Z) \end{equation}
 where
$*Z$ is the one-form  
\begin{equation}
*Z=k^\tau h_{\tau  \sigma} \, dx^\sigma
\end{equation}
This is a linearisation of the  Komar-type form for the NUT 4-momentum that was discussed in \cite{Bossard:2008sw}.

For the non-constant Killing vectors $k_\mu= \Lambda_{\mu\nu}x^\nu$
 (generating Lorentz transformations), the dual angular-momentum charge
 $Q[k]=  \frac 1 2 \tilde\Lambda_{\mu\nu} \tilde J^{\mu\nu}$
 can be written locally in terms of $\tilde h$ and this  is non-local when expressed in terms of 
   $h$ and $k_\mu$. An alternative local expression will be discussed in \cite{Hutt}.

%%%%%

\section {Kaluza Klein Monopoles and Other Solutions}
 \label {sols}

In this section some explicit solutions of linearised gravity will be considered and their electric and magnetic charges found.
 Of particular interest is the solution 
 that results from the linearisation of the Kaluza-Klein monopole solution of the full non-linear gravity or supergravity in $d$ dimensions. In the linear theory, this results from the superposition of a solution carrying mass and one carrying gravitational magnetic charge, so that these solutions can be considered separately.
It will be seen that, on dimensional reduction, the  magnetic gravitational charge $Q[\kappa]$ reduces to the magnetic charge of the graviphoton.

\subsection {Kaluza-Klein Ansatz}

The background spacetime in $d$ dimensions will be taken to be the product of 
 $d'=d-1$-dimensional Minkowski space with
 coordinates $x^m$ and a circle of radius $r$ with periodic coordinate $y$.
 On Kaluza Klein reduction on the circle  to $d'$ dimensions, $h_{\mu\nu}$ gives a $d'$-dimensional graviton $h_{mn}$, a $d'$-dimensional graviphoton $A_m=h_{my}$ and a scalar $\phi=h_{yy}$.
In this section, the solutions that will be discussed are ones in which all fields are independent of $y$.
(Note that one can also consider the $y$-independent system in $d$ dimensional Minkowski space with the $y$ direction non-compact, which can also be formulated in terms of the same fields in $d'$ dimensions.)

 In
 linearised gravity, one can consider solutions with, say, just $A_m=h_{my}$ non-zero. Then a solution $A_m$ of $d'$-dimensional Maxwell theory can be lifted to  a solution of $d$-dimensional linearised gravity and the electric and magnetic charges lift to gravitational charges in $d$ dimensions.
The electric charge $q$ in $d'$ dimensions lifts to the $y$-component $P^y$ of the momentum (\ref{momang}) as expected, while it will be shown that the magnetic charge $p_{m_1\dots m_{d'-4}}$ lifts to the component $\tilde P_{m_1\dots m_{d'-4}y\, |\, y}$ of the gravitational charge $Q[\kappa] $ defined by (\ref{pduaa}).

Consider the ansatz for the curvature
\begin{equation}
R_{m n p y} = \partial_p F_{m n}
\label {drgetre}
\end{equation}
If $\partial_{[p} F_{m n]}=0$, which implies that locally there is a 1-form potential $A$, with  $F=dA$, then  (\ref{drgetre}) is the linearised curvature (\ref{ris}) for 
\begin{equation}
h_{my}=2A_m
\end{equation}
  with $h_{mn}=0$, $h_{yy}=0$.
Then 
\begin{equation}  R_{m  y} = G_{m  y} =\partial^n F_{m n} \end{equation} 
so
\begin{equation}  T_{n y} = j_n\end{equation} 
The connection (\ref{conne}) has components
\begin{equation}  \Gamma_{m n\,|y } = \partial_{[m } h_{ n] y}
   = 2\partial_{[m  } A_{  n]} =F_{mn} 
   \label{gamans}
   \end{equation} 
\begin{equation}  \Gamma_{ n y \, |m } = %\frac{1}{2} 
 \partial_n A_m \end{equation}

If there are magnetic sources $\tilde j=*dF$, then
\begin{equation}  R_{[m n p] y} = \partial_{[p} F_{m n]}
=(*\tilde j)_{mnp} \end{equation} 
Magnetic monopoles in $d'$ dimensions have $\tilde j\ne 0$ at the locations of the monopoles and so give a violation of   the gravitational  Bianchi identity $R_{[m n p] y} =0$ in $d=d'+1$ dimensions.
The dual (\ref{sis}) of the curvature  is
\begin{equation}  S_{m_1 m_2 \ldots m_{d - 3} y | p y  } = \partial_p
   \widetilde{F_{}}_{m_1 m_2 \ldots .m_{d - 3}} \end{equation} 
with trace
 \begin{equation}  S'_{m_1 m_2 \ldots .m_{d - 4} y | y  } = \partial^p
   \widetilde{F_{}}_{m_1 m_2 \ldots m_{d - 4} p} \end{equation} 
   so that the dual gravitational source is
\begin{equation}   U_{m_1 m_2 \ldots m_{d - 2}  y| y  } =
   \tilde{j}_{m_2 \ldots .m_{d - 2}} \end{equation} 

The potential $A$ and the graviton $h$ are only defined in regions where $   \tilde{j}= 0$ so that $U= 0$.
Consider a  magnetic current $   \tilde{j}$  that is a delta-function so that the  source is a magnetic monopole
if $d'=d-1=4$ and is   a magnetic $d'-4$ brane if $d'>4$.
Removing the source location from the spacetime leaves a space ${\cal M}$ with non-trivial topology that can can support magnetic charge, with $A$ a gauge connection of a bundle on ${\cal M}$.

The Killing tensor given by (\ref{vkap}) with $V^\mu$ the unit vector in the $y$ direction, $V^\mu =(V^m,V^y)=(0,1)$, so that
\begin{equation}
\kappa   _{m_1\ldots m_{n-2}y  \, |\, y}= \alpha
  _{m_1\ldots m_{n-2} }
  \label{vkapaa}
\end{equation}
with $\alpha(x^m)$ a closed $n-2$ form.
Then the charge (\ref{asdert}),(\ref{pduaa}) is
\begin{equation}
Q[\kappa]=Q[V,\alpha]=(- )^{n}\frac 1 4  \frac 1 {d-1}\,
   \int _S \,  
\tilde \kappa    ^{ mn pq      \, |\,  y}
\Gamma _{  pq\, |y}
d \Sigma_{mn} 
\end{equation}
and using (\ref{gamans}) this gives
\begin{equation}
Q[V,\alpha]=(- )^{n} \frac 1 {d-1}\,
 \frac 1 4  \int _S \,  
(*\alpha)    ^{ mn pq   y  }
F_{  pq}
d \Sigma_{mn} 
\end{equation}
This is precisely the magnetic charge for Maxwell theory in $d'=d-1$ dimensions (\ref{qantib}) with
\begin{equation}
\tilde \sigma =(- )^{n} \frac 1 {d-1}\,
    \alpha
\end{equation}
For constant $\alpha$, (\ref{pduaa}) gives the charge $\tilde P ^{m_1\ldots m_{n-2}  y |y}$ with
\begin{equation}
Q[V,\alpha]
=\frac 1 {(n-2)!}\tilde P ^{m_1\ldots m_{n-2}  y |y}
 \alpha _{m _1\ldots m _{n-2} } 
 \label {pduaas}
\end{equation}
Choosing instead
\begin{equation}
\kappa   _{m_1\ldots m_{n-2}p  \, |\, q}=- \alpha
  _{[m_1\ldots m_{n-2} }\eta_{p]q}
  \label{vkapaaa}
\end{equation}
gives the same charge $Q[\kappa]$, a reflection of the fact that Killing tensors of the form (\ref{rhois}) give zero charge.

\subsection{Solutions with Electric and Magnetic Momentum}
\label {submag}

This and the following subsections will discuss solutions in $d=5$ dimensions
 in a background spacetime given by the product of a circle of radius $r$, with periodic coordinate $y\sim y+2\pi r$, and 4-dimensional Minkowski space,
with
coordinates    $x^m=(t,x^i)$ where $i=1,2,3$.
In this subsection, the only non-vanishing component of the graviton is $h_{my} $ giving a Maxwell field 
$h_{my}=A_m$ on reduction to 4 dimensions. The first  solution corresponds to a  set of electrically charged particles
in 4 dimensions with Coulomb potential $A_t=W(x)$   and the second 
  solution corresponds to a of a set of magnetically charged particles in 4 dimensions
for which the dual gauge field $\tilde A_m$ has a Coulomb potential $\tilde A_t=\tilde W$.

The electrically charged solution has
\begin{equation}
   h_{ty}=2A_t=W
 \end{equation}
with a Coulomb potential 
\begin{equation}
W(x^i)=-\sum_s \frac {q_s} {|x-x_s|}
\label {wiss}
\end{equation}
with sources at points $x_s^i$ in $\mathbb{R}^3$.
Dimensionally reducing on $y$ gives Maxwell theory in 4-dimensions with potential  $A_t=W$ and $A_i=0$,
which is the electric field for stationary electric charges $q_s$ at the points $x_s^i$ in $\mathbb{R}^3$. (Note that dimensional reduction gives a Maxwell action proportional to $2\pi r \int F\wedge *F$.)

The 4-d field strength is
$F_{it}=\partial _iW$ and the 5-d curvature is
\begin{equation}
R_{itj y} = \partial_i F_{jt}
=\partial _i\partial _j W
\end{equation}
so that
\begin{equation}  R_{ty} = \partial^i F_{it}
=  \Delta W
\end{equation} 
with 
$\Delta \equiv \partial _i\partial ^i $ and 
\begin{equation}\Delta W
=
4\pi \sum_s   {q_s} \delta (x-x_s)
\end{equation}
 This is a solution to the equations of motion with energy-momentum tensor 
 \begin{equation}
T_{ty}= 4\pi \sum_s   {q_s} \delta (x-x_s)
\end{equation}
Dimensionally reducing on the $y$ circle to 4 dimensions gives 
 an electric  current $j_m$
with $j_t=T_{ty},j_i=0$.

The charge $Q[k]$ with $k^\mu=\delta ^\mu_y$ the Killing vector generating translations in $y$ is given by the volume integral (\ref{qkis}) of (\ref{jtis}) with $\Sigma=\mathbb{R}^3\times S^1$, or by the surface integral (\ref{nester}) with $S=S^2\times S^1$ (with $S^2$ the sphere at infinity in $\mathbb{R}^3$)
giving the momentum $P^y$ (\ref{momang}) in the $y$ direction
\begin{equation}
P^y=Q[k]=8\pi^2 r  \sum_s   {q_s}
\end{equation}
 and agrees with the total electric charge in $d'=4$ dimensions.

There is a similar solution with magnetic charges given by
\begin{equation}
    h_{iy}=2A_i
    \label {hmagch}
 \end{equation}
with all other components vanishing. Here $A_i(x^j)$ is a vector potential on $\mathbb{R}^3$ with field strength
\begin{equation}
F_{ij}=\partial_i A_j-
  \partial_j A_i
\label{fij1}
\end{equation}
satisfying
\begin{equation}
F_{ij}=\epsilon _{ijk}\partial _k \widetilde W
\label {fij2}
\end{equation}
with
\begin{equation}
 \widetilde W(x^i)=- \sum_s \frac {p_s} {|x-x_s|}
\label{fij3}
\end{equation}
Dimensionally reducing on the $y$ circle to 4 dimensions gives 
 magnetic monopoles of charge $p_s$ at the points $x_s^i$ in $\mathbb{R}^3$.
For a single charge at $x_1=0$, in spherical polar coordinates the 2-form field strength
  is $F= p_1 \sin \theta d \theta \wedge d \phi$
and one can take $A=  p_1(1-\cos \theta)d\phi$.

The non-vanishing components of the curvature are
\begin{equation}
R_{ijk y} = \partial_k F_{ij}
\end{equation}
so that
\begin{equation}  R_{[ijk] y} = \partial_{[k} F_{ij]}
=\frac 1 3 \epsilon _{ijk} \Delta  \widetilde W
\end{equation} 
with \begin{equation}
\Delta  \tilde W
=4\pi
  \sum_s   {p_s} \delta (x-x_s)
\end{equation}
Then
the usual Bianchi identities $R_{[ijk] y} = \partial_{[k} F_{ij]}=0$ hold everywhere except at the locations $x_s$ of the monopoles.
This is a solution to the  equations of motion with dual source $U$ given by
\begin{equation}
U_{ty|y}=4\pi
\sum_s   {p_s} \delta (x-x_s)
\end{equation}
giving rise on dimensional reduction to a magnetic current
$\tilde j_m$ with $\tilde j_t=U_{ty|y}$.

In 5-dimensions, the Killing tensor $\kappa$ is a symmetric tensor $\kappa_{\mu\nu}=\kappa_{\nu\mu}$
satisfying
$\partial _{ [\rho  }\kappa_{\mu]\nu}=0$.
A non-trivial charge arises for the constant Killing tensor
$\kappa_{\mu\nu}=\delta _\mu^y
\delta _\nu^y
$. 
Note that this Killing tensor is closed but  not of the exact form (\ref{tyutr}): locally $\kappa_{yy}=\partial_y\partial_y\psi$
with $\psi  = \frac 1 2 y^2$ but this is not well-defined globally as $y$ is periodic.
The resulting charge $Q[\kappa]$ for this $\kappa$ is given by the volume integral (\ref{Qkkis}) of (\ref{kkis}), or the 
surface integral (\ref{Qkaph}), giving
\begin{equation}
 Q[\kappa]= 8\pi^2r \sum_s   {p_s}
 \label {KKcha}
\end{equation}
and agrees with the total magnetic charge in $d'=4$ dimensions.

\subsection{Solution with electric mass }

Again in the 5-dimensional set-up of the previous subsection, consider the ansatz
\begin{equation}
h_{tt}=aV, \qquad h_{yy}=bV, \qquad 
h_{ij}=cV\delta_{ij}
\end{equation}
for some function $V(x^i)$ and some constants $a,b,c$.
If
\begin{equation}
a=b+c
\label {asd}
\end{equation}
then the Ricci tensor is
\begin{equation}
R_{tt}=\frac a 2 \Delta V, \qquad R_{yy}=\frac b 2 \Delta V, \qquad 
R_{ij}=\frac c 2\Delta V\delta_{ij}
\end{equation}
and the Ricci scalar is
\begin{equation}
R=c\Delta V
\end{equation}
Taking
\begin{equation}
V(x^i)=- \sum_s \frac {m_s} {|x-x_s|}
\label{fij4}
\end{equation}
yields
\begin{equation}
\Delta V
=4\pi
  \sum_s   {m_s} \delta (x-x_s)
\end{equation}
This 
gives a solution with energy-momentum tensor
\begin{equation}
T_{tt}=\frac 1 2 (a+c)\Delta V, \qquad T_{yy}=\frac 1 2(b-c)\Delta V, \qquad 
T_{ij}=0\end{equation}

For a single-centre solution at $x_1=0$, if $b=0$ and $a=c$, the solution is a linearisation of the product of the 4-dimensional Schwarzschild solution, written in isotropic coordinates, and a circle, giving a black string  wrapping the $y$ circle. The dimensional reduction to $d'=4$ dimensions gives the linearised 4-dimensional Schwarzschild solution. If $a=0$ and $b=-1,c=1$ the solution is part of the linearisation of the Kaluza-Klein monopole solution, as will be discussed in  subsection \ref{KKMO}.

The mass $M$ is  given by the volume expression (\ref{jtis}),(\ref{qkis}) for $k^\mu =\delta ^\mu_t$:
\begin{equation}
M=\int d\Sigma _t\,  T_{tt}
\end{equation}
or by the surface integral (\ref{sadm}) which gives
\begin{equation}
M=-\frac 1 2 \int _S (h_{ij,j}-h_{jj,i}-h_{yy,i})
d\Sigma ^{0i}
\label{sadma}
\end{equation}
in agreement with \cite{Deser:1988fc}.
Both give the total mass as
\begin{equation}
M=4\pi^2 r(a+c)\sum_s  {m_s} 
\label {massic}
\end{equation}

\subsection{Solution with Magnetic Mass}
\label {magmass}

There is a similar solution to (\ref{hmagch}) with the role of $t$ and $y$ interchanged.
Taking
\begin{equation}
    h_{it}=2A'_i
    \label {magmag}
 \end{equation}
with all other components vanishing where as before $A'_i(x^j)$ is a vector potential on $\mathbb{R}^3$ with field strength 
\begin{equation}
F'_{ij}=\epsilon _{ijk}\partial _k X
\label{fij22}
\end{equation}
for a Coulomb potential
\begin{equation}
X(x^i)=- \sum_s \frac {n_s} {|x-x_s|}
\label{fij33}
\end{equation}
The non-vanishing components of the curvature are
\begin{equation}
R_{ijk t} = \partial_k F'_{ij}
\end{equation}
so that
\begin{equation}  R_{[ijk] t} = \partial_{[k} F'_{ij]}
=\frac 1 3 \epsilon _{ijk} \Delta  X
\end{equation} 
with \begin{equation}
\Delta  X
=
 4\pi \sum_s   {n_s} \delta (x-x_s)
\end{equation}
Then
the  Bianchi identities $R_{[ijk] t} = \partial_{[k} F_{ij]}=0$ hold everywhere except at the locations $x_a$.
This is a solution to the  equations of motion with dual source $U$ given by
\begin{equation}
 U_{yt\,  t}=4\pi
 \sum_s   {n_s} \delta (x-x_s)
\end{equation}

This solution carries the charge $Q[\lambda]$ with constant
$\lambda _{ty}=-\lambda _{yt}$
given by the volume integral (\ref{llis}),(\ref{Qllis}) or the surface integral (\ref{Qlamh}) both of which give
\begin{equation}
Q[\lambda]=8\pi^2 r \lambda _{ty} \sum _s n_s
\end{equation}
It also carries the charge
$Q[\kappa]$ with constant
 $\kappa _{ty}=\kappa _{yt}$
given by the 
 volume integral (\ref{kkis}),(\ref{Qkkis}) or the 
surface integral (\ref{Qkaph}), giving
\begin{equation}
 Q[\kappa]= 8\pi^2 r \kappa _{ty}\sum_s   {n_s}
\end{equation}

Dimensionally reducing on the $y$ circle  gives a solution $h_{mt}=A'_m$ of linearised gravity in 4-dimensional Minkowski space with
coordinates    $x^m=(t,x^i)$.
For a single source,
this
 can be viewed as a linearisation of the Lorentzian Taub-NUT solution \cite{Newman:1963yy,Taub} so that the 5-dimensional solution can be viewed as the product of linearised Lorentzian Taub-NUT with a circle.
 This  solution  was considered in \cite{Bunster:2006rt} where it was referred to as the gravitypole. 
The 4-dimensional linearised Taub-NUT space with $h_{it}=A'_i$  is a solution to the 4-dimensional equations of motion (\ref{4dein}) with dual source $\tilde T$  at $x=x_s$ given by
 $\tilde T_{t  t}=4\pi
  {n_s} \delta (x-x_s)$
while a linear superposition of such solutions has
\begin{equation}
 \tilde T_{t  t}=4\pi
 \sum_s   {n_s} \delta (x-x_s)
\end{equation}
For the time-translation Killing vector $k^\mu=\delta^\mu_t$, the charge 
is given by the volume integral
(\ref{jtisti}),(\ref{qkistilde}) 
or the surface integral (\ref{nesterdudu})
and is
\begin{equation}
\tilde Q[k] = 4\pi \sum_s  {n_s}
\label{sdfsdf}
\end{equation}

\subsection{Superpositions and the Linearized Kaluza-Klein Monopole}
\label{KKMO}

In the linearised gravity theory, linear combinations of solutions give new solutions. A linear superposition of the solutions of the previous subsections gives
\begin{equation}
   h_{ty}= W,\qquad  h_{iy}=2A_i,\qquad  h_{it}=2A'_i, \qquad 
   h_{tt}=aV, \qquad h_{yy}=bV, \qquad 
h_{ij}=cV\delta_{ij}
\label {supersol}
 \end{equation}
with
\begin{equation}
F'_{ij}=\epsilon _{ijk}\partial _k \widetilde W
,\qquad 
F'_{ij}=\epsilon _{ijk}\partial _k X
\label{fij22z}
\end{equation}
and
(\ref{wiss}),(\ref{fij3}),(\ref{fij4}),(\ref{fij33}) and (\ref{asd}).
There are then sources at the points $x_s^i$ with charges $(m_s,n_s,q_s,p_s)$.

A linearisation of the Kaluza-Klein monopole solution is obtained by taking 
$n_s=q_s=0$ and $m_s=p_s$ for each $s$ together with $a=0$ and $b=-1,c=1$
so that $V= \widetilde W$ and $W=X=0$.
This has mass (\ref{massic}) and the Kaluza-Klein monopole charge is given by (\ref{KKcha}).

\subsection{Global Structure}

A magnetic monopole of charge $p$ at the origin in 4 dimensions has
potential \begin{equation}
 \widetilde W(x^i)=-  p   \frac {1} {|x|}
\label{fij3x} 
\end{equation}
so that the
 field strength
(\ref{fij2})
 written in spherical polar coordinates  is the 2-form
\begin{equation}
F=  p  \sin \theta d\theta \wedge d \phi
\end{equation}
Two  choices of gauge potential 1-form are $A_\pm$ with
\begin{equation}
A_\pm=p(\pm 1-\cos \theta )  d \phi
\end{equation}
Here $A_+$ has a  Dirac string singularity at $\theta=\pi$ and
 $A_-$ has a  Dirac string singularity at $\theta=0$.
 If there is no charged matter coupling to $A$, then the Dirac strings are unobservable and either potential $A_\pm $ can be used.
 If there is charged matter, the Dirac strings are unobservable only if the electric charges all satisfy the Dirac quantisation condition.

The same results can also be understood in terms of the topology of the gauge bundle. Removing the origin from $\mathbb{R}^3$ leaves a topologicially non-trivial  space
 $(\mathbb{R}^3\setminus \{0\})\times \mathbb{R}\times S^1$ that can be covered by two patches,
 $H_\pm$. Here   $H_+$ is the region with $0\le \theta <\pi /2+\epsilon$   
  and $H_-$ is the region with
 $\pi/2- \epsilon< \theta \le\pi$. Then $A_\pm$ has no singularity in $H_\pm$ so $A_+$ can be used in $H_+$ and $A_-$ can be used in $H_-$.
 In the overlap, the potentials are related  by 
 \begin{equation}
A_+-A_-= 2 p d\phi 
\label {apm}
\end{equation}
In the absence of charged matter, the free Maxwell theory has a symmetry under shifts of the potential by a closed 1-form:
\begin{equation} 
A\to A +\alpha, \qquad d \alpha =0
\end{equation}
The two potentials $A_\pm $ differ by a closed form (\ref{apm}) and so in this case the gauge potentials are related by a gauge transformation of this type.
If there is charged matter, then the gauge symmetry is broken to shifts by \begin{equation}
A\to A + d \lambda
\end{equation}
The difference between the gauge potentials $A_\pm$ is then a gauge transformation with
\begin{equation}
\lambda = 2 p   \phi 
\end{equation}
which requires that $\lambda$ be a periodic variable with
$\lambda\sim \lambda + 4\pi p$, so that the gauge group is $U(1)$.
A charged field  $\psi$ of charge $q$ 
will transform under gauge transformations as $\psi\to e^{iq\lambda}\psi$.
The fields $\psi_\pm$ in $H_\pm$ are related by the gauge transformation
\begin{equation}
\psi_+=e^{iq\lambda}\psi_-
\end{equation}
with $\lambda = 2 p   \phi $, and this is well-defined provided
  that the charges satisfy the Dirac quantization condition
\begin{equation}
pq= \frac N 2
\label {dirac}
\end{equation}
for some integer $N$.

Now consider the magnetic solution of subsection \ref{submag} with $h_{iy}=A_i$
satisfying (\ref{fij1}),(\ref{fij2}) with a single centre potential (\ref{fij3x}).
Here $A$ can be taken to be $A_+$ or $A_-$, both of which have Dirac strings, giving rise to Dirac strings in the graviton field $h_{\mu\nu}$.
Electrically charged   particles in 4 dimensions lift to particles carrying momentum $P^y$ in the $y$ direction and if there are no particles carrying this momentum then the Dirac strings are unobservable.

Instead, removing the origin and taking the two patches $H_\pm$ with gravitons $h_\pm$
given by
\begin{equation}
(h_\pm)_{iy}=2(A_\pm)_i
\end{equation}
then $h_+ $ is non-singular in $H_+$ and   $h_- $ is non-singular in $H_-$. They are related by 
\begin{equation}
(h_+)_{iy}-(h_-)_{iy}=\partial _i\xi_y
\label {ertih}
\end{equation}
with
\begin{equation}
\xi_y=4 p\phi
\label {ertihw}
\end{equation}
If there is no  matter coupling to $h_{\mu\nu}$ then the theory is invariant under transformations
$h_{\mu\nu}\to h_{\mu\nu}+\delta h_{\mu\nu}$ where the linearised curvature $R_{\mu\nu\rho \sigma}(\delta h)$ of $\delta h_{\mu\nu}$ is zero and here the transition function (\ref{ertih}),(\ref{ertihw})
is such a symmetry transformation. Then the gauge fields in the two patches are related by such a transformation and so give a consistent configuration, without a Dirac quantization condition on the charges.

If there is charged matter, then the symmetry is (\ref{hvar}) and (\ref{ertih}),(\ref{ertihw}) is  a gauge transformation of this type provided
$\xi_y$ is a periodic parameter with $\xi_y\sim \xi_y+8 p\pi$.
This gives the surprising requirement that the gauge parameter $\xi_y$ should be periodic, although no such condition has arisen for the other components $\xi_t, \xi_i$. This can be understood by comparing with the full (non-linear) Kaluza-Klein monopole solution which has a circle fibration. The gauge transformation of the linear theory with parameter $\xi_y$ arises from  diffeomorphisms of  the circle fibres and it seems that the periodicity of the parameter $\xi_y$ arises from its origin as a circle diffeomorphism.
If there is matter with $y$-momentum $P^y=q$, then the electric and magnetic charges $p,q$ satisfy the quantization condition
(\ref{dirac}). (Note that one could also in principle allow the case in which $y$ is a coordinate for a non-compact dimension provided that $\xi_y$ remains periodic.)

Similar remarks apply to the solution with magnetic mass of subsection \ref{magmass} with $h_{it}=A_i
$. 
The global structure of this solution was also discussed in \cite{Hinterbichler:2022agn}.
Taking two patches with 
\begin{equation}
(h_\pm)_{it}=2(A_\pm)_i
\end{equation}
gives
\begin{equation}
(h_+)_{it}-(h_-)_{it}=\partial _i\xi_t
\label {ertiht}
\end{equation}
with
\begin{equation}
\xi_t=4 p\phi
\label {ertihwt}
\end{equation}
so that in this case it is $\xi_t$ that is periodic.
This reflects the necessity of taking time to be periodic in the non-linear Lorentzian Taub-NUT solution \cite{Newman:1963yy},\cite{Taub} and that $\xi_t$ arises from a diffeomorphism of the time circle in the non-linear solution. 

\subsection{Higher Dimensional Solutions}

Solutions in $d$ dimensions are obtained by taking the product of the 5-dimensional solutions (\ref{supersol}) given above with a $d-5$ torus. Let the coordinates on this torus be $u^a$
where $a=1,\dots,d-5$. 
Then for  solutions carrying the charges $p_s$, the 
source current $U_{\mu_{1}\mu_{2}\ldots\mu_{d-3}   \, |\,
 \nu }$ takes the form
\begin{equation}
U  _{a_1\ldots a_{d-5} y t \, |\, y}=4\pi\epsilon _{a_1\ldots a_{d-5} }
\sum_s   {p_s} \delta (x-x_s)
\end{equation}
Then a non-trivial charge is obtained by taking the Killing tensor in $d$ dimensions 
$\kappa  _{\mu_1\ldots \mu_{d-4}  \, |\, \rho}$   have non-vanishing components given by
\begin{equation}
\kappa  _{a_1\ldots a_{d-5} y \, |\, y}=\epsilon _{a_1\ldots a_{d-5} }
\end{equation}
which is the volume form on the $d-5$ torus. This is closed but not of the exact form (\ref{kaptriv}) and so gives a non-trivial charge $Q[\kappa]$.
This gives a bi-form charge $\tilde P _{\mu_1\ldots  \mu_{d-4}  |\rho}
$ given by (\ref{pduaax})
with components 
\begin{equation}
\tilde P _{a_1\ldots  a_{d-5} y |y}
\propto
\epsilon _{a_1\ldots a_{d-5} }
\end{equation}

For the 4-dimensional solution (\ref{magmag}),(\ref{fij22}),(\ref{fij33}) with magnetic mass carrying charges $n_s$ one can take the product with a $d-4 $ torus with coordinates   $v^\alpha $
where $\alpha=1,\dots,d-4$. 
The current is now
\begin{equation}
 U_{{\alpha_1\ldots \alpha_{d-4} }t\, | t}=4\pi \epsilon _{\alpha_1\ldots \alpha_{d-4} }
 \sum_s  {n_s} \delta (x-x_s)
\end{equation}
In this case one can take the $d$-dimensional Killing tensors 
$\lambda  _{\mu_1\ldots \mu_{d-3}}$ and $\kappa  _{\mu_1\ldots \mu_{d-4}  \, |\, \rho}$ to be
\begin{equation}
\kappa  _{
\alpha_1\ldots \alpha_{d-4}  \, |\, t}=\epsilon _{\alpha_1\ldots \alpha_{d-4} }
\end{equation}
and
\begin{equation}
\lambda  _{\alpha _1\ldots \alpha _{d-4}  t}=\epsilon _{\alpha_1\ldots \alpha_{d-4} }
\end{equation}
This then leads to charges $Q[\lambda]$ and $Q[\kappa]$ giving   charges
$\hat P _ {\mu_1\ldots \mu_{d-3}}$ given by (\ref{maghat})
and $\tilde P _{\mu_1\ldots  \mu_{d-4}  |\rho}
$ given by (\ref{pduaax}) with
\begin{equation}
\hat P _ {\alpha_1\ldots \alpha_{d-4}t}
\propto
\epsilon _{\alpha_1\ldots \alpha_{d-4} }
, \qquad
\tilde P _{\alpha_1\ldots  \alpha_{d-4}  |t}
\propto
\epsilon _{\alpha_1\ldots \alpha_{d-4} }
\end{equation}

\section {Discussion} \label{discu}

Invariances of a gauge theory -- i.e.\ gauge transformations that leave the gauge field (plus any other fields in the theory) invariant -- can be regarded as global symmetries and have corresponding Noether charges. For free theories, this gives a rich set of charges as there are invariances for {\it any} field configuration. For interacting theories, however, there are typically no such invariances as invariances only arise for special field configurations. For example, for general relativity the invariances are isometries of the spacetime, and most spacetimes do not have any isometries. Such charges are then only defined for those spacetimes with isometries. 
Fortunately, the  construction generalises to spaces with asymptotic isometries:
for asymptotically flat or asymptotically AdS spacetimes, 
if suitable boundary conditions are satisfied, then an asymptotic invariance is sufficient for there to be conserved ADM charges \cite{Abbott:1981ff}. For the free theories with exact invariances there is a  charge conservation and charges can be defined as an integral over any closed $d-2$ surface. On the other hand,    for the interacting theories with asymptotic invariances  the surface must be a sphere at infinity and there is no local charge -- e.g.\ for linearised gravity there is a local concept of energy but for general relativity energy is non-local and the ADM mass is defined by a sphere at infinity.  Here, attention has been restricted to exact invariances and the case of asymptotic invariances will be discussed elsewhere.

The discussion here extends readily to  tensor fields with   Young tableaux with $N$ columns. For a gauge field in such a representation  there are $N$ independent symmetries if the all the columns have different lengths (if some of the lengths are equal, there will be less independent symmetries) \cite{Curtright:1980yk}-\cite{Dubois-Violette:1999rd},\cite{deMedeiros:2002qpr} and for each symmetry  a charge can be defined in the way discussed here. The differential calculus of multiforms introduced in \cite{deMedeiros:2002qpr,Bekaert:2002dt} provides an elegant way of formulating the theory.

The dual formulation of the theory  will have its own electric type (Noether) symmetries and these in turn will have associated conserved charges. The electric-type charges of the dual theory  can then be interpreted as magnetic-type charges for the original theory.
This has been seen in detail for the free graviton and dual graviton theories but the method applies more generally.

In 4 dimensions the electric gravitational charges are the momentum $P^\mu$ and the angular momentum $J^{\mu\nu}$ while the magnetic ones discussed in section
\ref{4DSec}
are the dual
momentum $\tilde P^\mu$ and   angular momentum $\tilde J^{\mu\nu}$.
The expression found for the angular momentum $J^{\mu\nu}$ depends explicitly on the   graviton field $h_{\mu\nu}$ and so cannot be written locally for the dual graviton theory, so that it can be thought of as a non-local charge there.
Similarly, the dual
angular momentum $\tilde J^{\mu\nu}$ depends explicitly on the  dual  graviton field $\tilde h_{\mu\nu}$ and so cannot be written locally for the   graviton theory, so that it can be viewed as a non-local charge for linearised gravity.
The linearised Lorentzian Taub-NUT solution carries the dual mass $\tilde P^0$, often referred to as the   NUT charge.

In $d>4 $ dimensions
the electric gravitational charges are again the momentum $P^\mu$ and the angular momentum $J^{\mu\nu}$ while the magnetic ones from $Q[\lambda]$ are
the $d-3$-form charge $\hat P _ {\mu_1\ldots \mu_{d-3}}$ and the  $d-2$-form charge  $\hat J _{\mu_1\ldots \mu_{d-2} }$
given by (\ref{maghat})
while for constant $\kappa$ there is a magnetic $[d-4,1]$ bi-form charge $\tilde P _{\mu_1\ldots  \mu_{d-4}  |\rho}
$ given by (\ref{pduaax}).
As was seen in section \ref{sols}, the
 Kaluza-Klein monopole with Kaluza-Klein direction given by a coordinate $y$
carries the $d-5$ form charge
$\tilde P _{i_1\ldots  i_{d-5} y |y}
$ that is a spatial $d-5$ form orthogonal to the $y$ direction 
(where the coordinates $x^i$ are the spatial coordinates orthogonal to   $y$). 
The product of the
linearised Lorentzian Taub-NUT solution with a $d-4$ dimensional torus carries
the charges
carries the charges 
$\tilde P _{\alpha_1\ldots  \alpha_{d-4}  |t}
$ and $\hat P _ {\alpha_1\ldots \alpha_{d-4}t}$, both of which are spatial $d-4$ forms that are
proportional to the volume form on the internal torus.

 The expression (\ref{nester}) for the charge charge $P^m$ is the linearisation of Nester's expression for the ADM momentum given in \cite{Nester:1981bjx}, while the expression
 obtained here for $\tilde P _{i_1\ldots  i_{d-5} y |y}
$ agrees with the linearisation of the magnetic $d-5$ form charge for gravity introduced in \cite{Hull:1997kt}. For example, for $d=11$ this is the 6-form charge occurring in the $d=11$ supersymmetry algebra that is carried by Kaluza-Klein monopoles. It will be interesting to understand better the roles of the other charges introduced here for the linearised theory.

These charges for the full non-linear supergravity theory can be written as follows.
For a  metric $g_{\mu\nu}$ approaching a  background metric $\bar g_{\mu\nu}$ asymptotically with corresponding   vielbeins $  e_\mu ^a, \bar e_\mu ^a$ and spin-connections
$\omega _\mu {}^a{}_b (e)$ and $\bar \omega _\mu {}^a{}_b (\bar e)$, the differences between the connections is asymptotically a tensor
\begin{equation}
\Gamma_\mu {}^a{}_b =\omega _\mu {}^a{}_b (e)-\bar \omega _\mu {}^a{}_b (\bar e)
\end{equation}
Here we will take the background metric $\bar g_{\mu\nu}$ to be that of Minkowski space. Note that we only require that the spacetime $(M, g_{\mu\nu})$ have an asymptotic region that
approaches the background and the geometry and topology of $M$ can be very different from that of Minkoswki space outside this asymptotic region.
Then the expression for the ADM momentum $P^m=Q[k^m]$ given in \cite{Nester:1981bjx}
is
\begin{equation}Q[k]= \frac 1 4 \int_S \delta ^{\sigma \tau \lambda}_{\mu\nu \rho}
 \Gamma ^{\nu a}{}_b k^{ \rho} 
 \bar e^\mu_a \bar e ^b_\lambda  d \Sigma_{\sigma \tau}
\end{equation}
and this indeed reduces to (\ref{nester}) for the linearised theory.
In terms of 1-forms
\begin{equation} 
e^a=e^a_\mu dx^\mu, \qquad k= k_\mu dx^\mu,\qquad 
\Gamma^a{}_b=\Gamma_\mu {}^a{}_b dx^\mu
\end{equation}
this can be written as  \cite{Hull:1983ap}
\begin{equation}Q[k]=\frac 1 4 \int_S * \Gamma_{ab} \wedge k\wedge \bar e^a\wedge \bar e ^b
\end{equation}
with
\begin{equation}\Gamma =\frac 1 2 \Gamma_{ab} \wedge   \bar e^a\wedge \bar e ^b
\end{equation}
The $K$-charge corresponding to  a $d-5$ form  $\rho$ 
is given  by \cite{Hull:1997kt}
\begin{equation} 
Q[\rho]=\frac 1 4 \int_S  \rho \wedge \Gamma_{ab} \wedge   \bar e^a\wedge \bar e ^b
\label {kcharge}
\end{equation}
%This can then be written in the form (\ref{Kwedge}) 
This suggests the interesting question of whether   the other charges introduced here have extensions to the full non-linear theory of gravity: this will be addressed elsewhere.

In the non-linear theory, gravitational energy is non-local, so that only the total energy in the spacetime can be defined. Moreover, introducing magnetic charges changes the topology of spacetime (or introduces Dirac string singularities) and the non-trivial topology leads to some issues in the definition of the   charges, as discussed in e.g.\  \cite{Hull:1997kt,Ramaswamy,Ashtekar,Deser:1988fc,Bombelli:1986sb}.
The linear theory has the advantage that energy is local and the charge contained within any surface $S$ can be defined.
In the linear theory there is no longer a relation between the graviton field and the geometry and topology of spacetime and one can work in a flat space background, so that the nature of the magnetic charges is clearer than it is in the full interacting theory.

An alternative approach to constructing  charges for a given theory from the conserved charges of a dual formulation of that theory was given in \cite{Nguyen:2022nnx} and in particular an infinite number of charges were found there.
 In general, a theory has many dual formulations \cite{Hull:2001iu,Boulanger:2015mka}. For example, a $p$-form gauge field has a dual  formulation in terms of a $d-p-2$ form gauge field together with \lq exotic' duals consisting of gauge fields with 
 $[d-2,d-2,\dots ,p]$ and $[d-2,d-2,\dots ,d-p-2]$ Young tableaux  (where the lengths of the columns are listed in square brackets) with an arbitrary number of columns of length $d-2$. Conserved charges associated with the gauge invariances of examples of such exotic duals were constructed in \cite{Nguyen:2022nnx}.
 There is a parent action that gives rise to both the graviton and dual graviton theories \cite{West:2001as,West:2014qoa,Tumanov:2017whf,Boulanger:2008nd,Hohm:2018qhd} which has a certain shift symmetry and the charge associated with this was also constructed in \cite{Nguyen:2022nnx}.

There are alternative covariant forms for the ADM charges in the linear and non-linear theories \cite{Penrose:1982wp} and \cite{Hinterbichler:2022agn,Benedetti:2021lxj,Benedetti:2023ipt,Gomez-Fayren:2023qly}. The extension of these to covariant forms for the magnetic charges found here will be discussed in \cite{Hutt}.
In \cite{Godazgar:2018qpq}, a  BMS-type generalisation of the dual momentum in four dimensions 
of \cite{Ramaswamy,Ashtekar}
was found and it would be interesting to consider similar generalisations of the charges discussed here.

\section*{Acknowledgements}

This  work   was  supported by 
 the STFC Consolidated Grant ST/T000791/1. I would like to thank Ulf Lindstrom and Max Hutt for helpful discussions.

\appendix
\section{Surface Integral Form of Charges}

 \subsection{The dual charges $Q[\lambda]$}

The charge $Q[\lambda]$ is given by  (\ref{Qllis}).
The current (\ref{llis}) is 
\begin{equation} 
j_\nu [\lambda ]= 
\frac 1 {n!} \lambda  _ {\mu_1\ldots \mu_n} E^{ \mu_{1}\mu_{2}\ldots\mu_{n}   \, |\, \nu }
\label{llisas}
\end{equation}
using the field equation (\ref{duein}).
If there are no electric sources, $T_{\mu\nu}=0$, then the dual field strength $S$ is given by (\ref{dufistr})
and $E$ is given by (\ref{seindell}) so that
\begin{eqnarray}
j^\nu [\lambda ]
%&=& \frac 1 {n!} \lambda  _ {\mu_1\ldots \mu_n} E^{ \mu_{1}\mu_{2}\ldots\mu_{n}   \, |\, \nu } \\
&=& 
\frac 1 {n!} \lambda  _ {\mu_1\ldots \mu_n}    \partial _\rho   \partial _\sigma  L ^{\mu_{1}\mu_{2}\ldots\mu_{n} \rho} {}^{| \nu \sigma}
\nonumber
\\
&=& 
\frac 1 {n!}
  \partial _\rho  \left[
 \lambda  _ {\mu_1\ldots \mu_n}    \partial _\sigma  L ^{\mu_{1}\mu_{2}\ldots\mu_{n} \rho} {}^{| \nu \sigma}
 -
 L ^{\mu_{1}\mu_{2}\ldots\mu_{n} \sigma} {}^{| \nu \rho}   \partial _\sigma \lambda  _ {\mu_1\ldots \mu_n} 
 \right]
\end{eqnarray}
using the fact that the Killing-Yano tensors (\ref{lkilt}) satisfy  
\begin{equation}  \partial_\rho    \partial _\sigma  \lambda  _ {\mu_1\ldots \mu_n} 
=0
\label{ddlam}
\end{equation}
Let $Y^{\nu\rho}=-Y^{\rho\nu}$ be the antisymmetric tensor
\begin{equation}
Y^{\nu\rho}=\frac 1 {n!}
 \lambda  _ {\mu_1\ldots \mu_n}  \left[
  \partial _\sigma  L ^{\mu_{1}\mu_{2}\ldots\mu_{n} \rho} {}^{| \nu \sigma}
 -
   \partial _\sigma  L ^{\mu_{1}\mu_{2}\ldots\mu_{n} \nu} {}^{|  \rho \sigma}
  \right]
 -\frac 1 {n!}
 L ^{\mu_{1}\mu_{2}\ldots\mu_{n} \sigma} {}^{| \nu \rho}   \partial _\sigma \lambda  _ {\mu_1\ldots \mu_n} 
\end{equation}
so that
\begin{equation}
j^\nu [\lambda ]=   \partial _\rho Y^{\nu\rho} +Z^\nu
\end{equation}
where
\begin{eqnarray}
Z^\nu  &=&\frac 1 {n!}
 \partial _\rho \left[
  \lambda  _ {\mu_1\ldots \mu_n}    \partial _\sigma  L ^{\mu_{1}\mu_{2}\ldots\mu_{n} \nu} {}^{|  \rho \sigma}
  \right]
  \nonumber
  \\
   &=&\frac 1 {n!}
(\partial _\rho 
  \lambda  _ {\mu_1\ldots \mu_n} )   \partial _\sigma  L ^{\mu_{1}\mu_{2}\ldots\mu_{n} \nu} {}^{|  \rho \sigma}
\nonumber
  \\
  &=&\frac 1 {n!}
 \partial _\sigma \left[
 (\partial _\rho \lambda  _ {\mu_1\ldots \mu_n}   )   L ^{\mu_{1}\mu_{2}\ldots\mu_{n} \nu} {}^{|  \rho \sigma}
  \right]
  \nonumber
   \\
  &=&
 - \frac 1 {n!}\partial _\rho \left[
 (\partial _  \sigma \lambda  _ {\mu_1\ldots \mu_n}   )   L ^{\mu_{1}\mu_{2}\ldots\mu_{n} \nu} {}^{|  \rho \sigma}
  \right]
   \nonumber
   \\
  &=&
-\frac 1 {n!} \partial _\rho \left[
 (\partial _  \sigma \lambda  _ {\mu_1\ldots \mu_n}   )   \eta^{\lambda \mu_{1}\mu_{2}\ldots\mu_{n}  \nu  } {}^ {|  \sigma   \rho \tau_{1}\tau_{2}\ldots \tau_{n}  } \,\,
 D_{\tau_{1}\ldots \tau_{n}  \, |\,\lambda } 
  \right]
\end{eqnarray}
Using (\ref{lkill}), $ \eta^{[\lambda \mu_{1}\mu_{2}\ldots\mu_{n}  \nu  } {}^ {|  \sigma ]  \rho \tau_{1}\tau_{2}\ldots \tau_{n}  }
=0$ and $D_{[\tau_{1}\ldots \tau_{n}  \, |\,\lambda ]}=0$
this can be rewritten as
\begin{eqnarray}
Z^\nu  &=&-\frac 1 {n+1}\frac 1 {n!}
 \partial _\rho \left[
 (\partial _  \sigma \lambda  _ {\mu_1\ldots \mu_n}   )   \eta^{\lambda \mu_{1}\mu_{2}\ldots\mu_{n}  \sigma  } {}^ {|    
 \nu \rho \tau_{1}\tau_{2}\ldots \tau_{n}  } \,\,
 D_{\tau_{1}\ldots \tau_{n}  \, |\,\lambda } 
  \right]
   \nonumber
   \\
  &=&
  -\frac 1 {n+1}\frac 1 {n!}
   \partial _\rho \left[
 (\partial _  \sigma \lambda  _ {\mu_1\ldots \mu_n}   )  
  L ^{\mu_{1}\mu_{2}\ldots\mu_{n} \sigma} {}^{|  \rho  \nu}
  \right]
\end{eqnarray}
Then 
$$j_\mu[\lambda] =   \partial ^\nu J_{\mu \nu}[\lambda]$$
with
\begin{eqnarray}
J^{\nu\rho}[\lambda]&=&Y^{\nu\rho}-\frac 1 {n+1}\frac 1 {n!}
    (\partial _  \sigma \lambda  _ {\mu_1\ldots \mu_n}   )  
  L ^{\mu_{1}\mu_{2}\ldots\mu_{n} \sigma} {}^{|  \rho  \nu}
 \end{eqnarray}
 so that
 \begin{equation}
J^{\nu\rho}=
\lambda  _ {\mu_1\ldots \mu_n}  \left[
    \partial _\sigma  L ^{\mu_{1}\mu_{2}\ldots\mu_{n} \rho} {}^{| \nu \sigma}
 -
      \partial _\sigma  L ^{\mu_{1}\mu_{2}\ldots\mu_{n} \nu} {}^{|  \rho \sigma}
  \right]
 - \frac n {n+1}
 L ^{\mu_{1}\mu_{2}\ldots\mu_{n} \sigma} {}^{| \nu \rho}   \partial _\sigma \lambda  _ {\mu_1\ldots \mu_n} 
\end{equation}
Using (\ref{seindellaaa}) this can  be written as 
\begin{eqnarray}
J^{\nu\rho}[\lambda]&=&\frac 1 {n!}
\lambda  _ {\mu_1\ldots \mu_n}  \left[
     \eta^{\lambda \mu_{1}\mu_{2}\ldots\mu_{n}  \rho  } {}^ {|   \tau_{1}\tau_{2}\ldots \tau_{n} \sigma\nu} \,\,
  -
    \eta^{\lambda \mu_{1}\mu_{2}\ldots\mu_{n}     \nu } {}^ {|   \tau_{1}\tau_{2}\ldots \tau_{n} \sigma \rho} \,\,
  \right] \tilde \Gamma _{\sigma\tau_{1}\ldots \tau_{n}  \, |\,\lambda } 
  \nonumber
   \\
   &+& \frac n {n+1} \frac 1 {n!}
% L ^{\mu_{1}\mu_{2}\ldots\mu_{n} \sigma} {}^{| \nu \rho}  
 \partial _\sigma \lambda  _ {\mu_1\ldots \mu_n} 
  \eta^{\lambda \mu_{1}\mu_{2}\ldots\mu_{n} \sigma   } {}^ {| \nu    \rho \tau_{1}\tau_{2}\ldots \tau_{n}  } \,\,
 D_{\tau_{1}\ldots \tau_{n}  \, |\,\lambda }
\end{eqnarray}
For constant $\lambda$ the second line of this equation  vanishes
and $J$ can be rewritten as
\begin{equation}
J^{\nu\rho}[\lambda]=  (-1)^{n+1}\frac {1} {(n+2)}  \tilde \lambda^{  \alpha \beta\rho} (*\tilde \Gamma )_{\alpha }{}^\nu {}_{|\beta }-(\nu \leftrightarrow \rho)
\end{equation}
If the \lq torsion' $(*\tilde \Gamma )_{[\alpha \nu \, |\lambda ] } =0$ this can be rewritten as
\begin{equation}
J^{\nu\rho}[\lambda]= (-1)^{n+1}
 \frac {1} {2(n+2)} 
 \tilde \lambda^{  \alpha \beta\rho} 
 (*\tilde \Gamma )_{\alpha\beta  }{}^\nu
 -(\nu \leftrightarrow \rho)
\label{jlambloca}
\end{equation}

This current  can alternatively be written in terms of $\hat \Gamma$ instead of $\tilde \Gamma$.
 First, the current can be written as
\begin{eqnarray}
j_\nu [\lambda ]
%&=& \frac 1 {n!} \lambda  _ {\mu_1\ldots \mu_n} E^{ \mu_{1}\mu_{2}\ldots\mu_{n}   \, |\, \nu } \\
&=& 
\frac 1 {n!} \lambda  _ {\mu_1\ldots \mu_n}    \partial _\rho   \partial _\sigma  L ^{\mu_{1}\mu_{2}\ldots\mu_{n} \rho} {}^{| \nu \sigma}
\nonumber
\\
&=& 
\frac 1 {n!}
  \partial _\sigma  \left[
 \lambda  _ {\mu_1\ldots \mu_n}    \partial _\rho  L ^{\mu_{1}\mu_{2}\ldots\mu_{n} \rho} {}^{| \nu \sigma}
 -
 L ^{\mu_{1}\mu_{2}\ldots\mu_{n} \sigma} {}^{| \nu \rho}   \partial _\rho \lambda  _ {\mu_1\ldots \mu_n} 
 \right]
\end{eqnarray}
using (\ref{ddlam}).
As a result, a
    2-form current  $J$ with
$$j_\mu[\lambda] =   \partial ^\nu J_{\mu \nu}[\lambda]$$
is
\begin{equation}
J^ {\sigma \nu}[\lambda]=
\frac 1 {n!}
  \left[
 \lambda  _ {\mu_1\ldots \mu_n}    \partial _\rho  L ^{\mu_{1}\mu_{2}\ldots\mu_{n} \rho} {}^{| \nu \sigma}
 -
 L ^{\mu_{1}\mu_{2}\ldots\mu_{n} \sigma} {}^{| \nu \rho}   \partial _\rho \lambda  _ {\mu_1\ldots \mu_n} 
 \right]
\end{equation}
with the   charge  given by the surface integral
\begin{equation}
 Q[\lambda]=
\frac 1 2  \int _S \, J _{\mu \nu} [\lambda]\, d \Sigma^{\mu \nu}
\end{equation}
For constant Killing-Yano tensors $\lambda$, the current reduces to
\begin{equation}
J^ {\sigma \nu}[\lambda]=
\frac 1 {n!}
 \lambda  _ {\mu_1\ldots \mu_n}    \partial _\rho  L ^{\mu_{1}\mu_{2}\ldots\mu_{n} \rho} {}^{| \nu \sigma}
\end{equation}
Note that 
\begin{equation}
  \partial _\rho  L ^{\mu_{1}\mu_{2}\ldots\mu_{n} \rho} {}^{| \nu \sigma}=  
     \eta^{\lambda\mu_{1}\mu_{2}\ldots\mu_{n}  \rho  } {}^ {|   \tau_{1}\tau_{2}\ldots \tau_{n} \sigma\nu} \,\,
\hat  \Gamma
 _{\tau_{1}\ldots \tau_{n}  \, |\,\lambda \rho } 
\end{equation}
where $ \hat  \Gamma _{\mu_{1}\mu_{3}\ldots\mu_{n}} {}
_{\sigma\tau }$ is the connection (\ref{hatducon}), so that
\begin{equation}
J^ {\sigma \nu}[\lambda]=\frac
1 {n!}
 \lambda  _ {\mu_1\ldots \mu_n}    \,
 \eta^{\mu_{1}\mu_{2}\ldots\mu_{n} \lambda \rho  } {}^ {|   \tau_{1}\tau_{2}\ldots \tau_{n} \sigma\nu} \,\,
\hat  \Gamma
 _{\tau_{1}\ldots \tau_{n}  \, |\,\lambda \rho } 
\end{equation}
which can be rewritten as 
\begin{equation}
J_ {\sigma \nu}[\lambda]=
- \frac 1 {(n+1)(n+2)}\,
\tilde  \lambda  ^{\lambda \rho  \mu} 
(*\hat  \Gamma)
_
{ \sigma\nu \mu \, | \, \lambda \rho } 
\end{equation}
where
\begin{equation}
(*\hat  \Gamma)
^
{ \sigma\nu \mu \, |}{}_{ \, \lambda \rho } =\frac 1 {n!}\epsilon ^ {   \tau_{1}\tau_{2}\ldots \tau_{n} \sigma\nu \mu} \hat  \Gamma
 _{\tau_{1}\ldots \tau_{n}  \, |\,\lambda \rho } 
\end{equation}


\begin{thebibliography}{99}

%\cite{Hull:2000zn}
\bibitem{Hull:2000zn}
C.~M.~Hull,
``Strongly coupled gravity and duality,''
Nucl.\ Phys.\ B {\bf 583} (2000) 237
[arXiv:hep-th/0004195].
%%CITATION = HEP-TH 0004195;%%
%\cite{Hull:2001iu}

%\cite{Hull:1994ys}
\bibitem{Hull:1994ys}
C.~M.~Hull and P.~K.~Townsend,
``Unity of superstring dualities,''
Nucl. Phys. B \textbf{438} (1995), 109-137
%doi:10.1016/0550-3213(94)00559-W
[arXiv:hep-th/9410167 [hep-th]].
%1748 citations counted in INSPIRE as of 13 Aug 2023



%\cite{Hull:1997kt}
\bibitem{Hull:1997kt}
C.~M.~Hull,
``Gravitational duality, branes and charges,''
Nucl. Phys. B \textbf{509} (1998), 216-251
%doi:10.1016/S0550-3213(97)00501-4
[arXiv:hep-th/9705162 [hep-th]].
%176 citations counted in INSPIRE as of 25 Aug 2022



%\cite{Abbott:1981ff}
\bibitem{Abbott:1981ff}
L.~F.~Abbott and S.~Deser,
``Stability of Gravity with a Cosmological Constant,''
Nucl. Phys. B \textbf{195} (1982), 76-96
%doi:10.1016/0550-3213(82)90049-9
%924 citations counted in INSPIRE as of 23 Aug 2022



%\cite{Hull:2001iu}
\bibitem{Hull:2001iu}
  C.~M.~Hull,
  ``Duality in gravity and higher spin gauge fields,''
  JHEP {\bf 0109} (2001) 027
  %doi:10.1088/1126-6708/2001/09/027
  [hep-th/0107149].
  %%CITATION = doi:10.1088/1126-6708/2001/09/027;%%
  %112 citations counted in INSPIRE as of 08 Dec 2017




%\cite{Hull:2000rr}
\bibitem{Hull:2000rr}
  C.~M.~Hull,
  ``Symmetries and compactifications of (4,0) conformal gravity,''
  JHEP {\bf 0012} (2000) 007
  %doi:10.1088/1126-6708/2000/12/007
  [hep-th/0011215].
  %%CITATION = doi:10.1088/1126-6708/2000/12/007;%%
  %52 citations counted in INSPIRE as of 08 Dec 2017

%\cite{Hull:2000ih}
\bibitem{Hull:2000ih}
  C.~M.~Hull,
  ``Conformal nongeometric gravity in six-dimensions and M theory above the Planck energy,''
  Class.\ Quant.\ Grav.\  {\bf 18} (2001) 3233
  %doi:10.1088/0264-9381/18/16/313
  [hep-th/0011171].
  %%CITATION = doi:10.1088/0264-9381/18/16/313;%%
  %15 citations counted in INSPIRE as of 08 Dec 2017

\bibitem{West:2001as}  P.~C.~West,
 ``$E_{11}$ and M theory",
  Class.\ Quant.\ Grav.\  {\bf 18} (2001) 4443
  [arXiv:hep-th/0104081].
  %%CITATION = CQGRD,18,4443;%%



%\cite{deMedeiros:2002qpr}
\bibitem{deMedeiros:2002qpr}
  P.~de Medeiros and C.~Hull,
  ``Exotic tensor gauge theory and duality,''
  Commun.\ Math.\ Phys.\  {\bf 235} (2003) 255
  %doi:10.1007/s00220-003-0810-z
  [hep-th/0208155].
  %%CITATION = doi:10.1007/s00220-003-0810-z;%%
  %88 citations counted in INSPIRE as of 08 Dec 2017

  %\cite{deMedeiros:2003osq}
\bibitem{deMedeiros:2003osq}
  P.~de Medeiros and C.~Hull,
  ``Geometric second order field equations for general tensor gauge fields,''
  JHEP {\bf 0305} (2003) 019
  %doi:10.1088/1126-6708/2003/05/019
  [hep-th/0303036].
  %%CITATION = doi:10.1088/1126-6708/2003/05/019;%%
  %75 citations counted in INSPIRE as of 08 Dec 2017


\bibitem{Henneaux:2004jw}  M.~Henneaux and C.~Teitelboim,
  {\sl Duality in linearized gravity},
  Phys.\ Rev.\  D {\bf 71} (2005) 024018
  [arXiv:gr-qc/0408101].
  %%CITATION = PHRVA,D71,024018;%%

%\cite{Bunster:2006rt}
\bibitem{Bunster:2006rt}
C.~W.~Bunster, S.~Cnockaert, M.~Henneaux and R.~Portugues,
``Monopoles for gravitation and for higher spin fields,''
Phys. Rev. D \textbf{73} (2006), 105014
%doi:10.1103/PhysRevD.73.105014
[arXiv:hep-th/0601222 [hep-th]].
%49 citations counted in INSPIRE as of 25 Aug 2022

  %\cite{Bunster:2013oaa}
\bibitem{Bunster:2013oaa}
  C.~Bunster, M.~Henneaux and S.~Hortner,
  ``Twisted Self-Duality for Linearized Gravity in D dimensions,''
  Phys.\ Rev.\ D {\bf 88} (2013) no.6,  064032
  %doi:10.1103/PhysRevD.88.064032
  [arXiv:1306.1092 [hep-th]].
  %%CITATION = doi:10.1103/PhysRevD.88.064032;%%
  %10 citations counted in INSPIRE as of 08 Dec 2017
  
  



%\cite{West:2014qoa}
\bibitem{West:2014qoa}
  P.~West,
  ``Dual gravity and E11,''
  arXiv:1411.0920 [hep-th].
  %%CITATION = ARXIV:1411.0920;%%
  %5 citations counted in INSPIRE as of 08 Dec 2017

%%%%%
%\cite{Tumanov:2017whf}
\bibitem{Tumanov:2017whf}
  A.~G.~Tumanov and P.~West,
  ``E11 and the non-linear dual graviton,''
  arXiv:1710.11031 [hep-th].
  %%CITATION = ARXIV:1710.11031;%%
  
  
%\cite{Boulanger:2008nd}
\bibitem{Boulanger:2008nd}
N.~Boulanger and O.~Hohm,
``Non-linear parent action and dual gravity,''
Phys. Rev. D \textbf{78} (2008), 064027
%doi:10.1103/PhysRevD.78.064027
[arXiv:0806.2775 [hep-th]].
%42 citations counted in INSPIRE as of 25 Oct 2023

%\cite{Hohm:2018qhd}
\bibitem{Hohm:2018qhd}
O.~Hohm and H.~Samtleben,
``The dual graviton in duality covariant theories,''
Fortsch. Phys. \textbf{67} (2019) no.5, 1900021
%doi:10.1002/prop.201900021
[arXiv:1807.07150 [hep-th]].
%6 citations counted in INSPIRE as of 25 Oct 2023
  
%\cite{Bekaert:2002dt}
\bibitem{Bekaert:2002dt}
X.~Bekaert and N.~Boulanger,
``Tensor gauge fields in arbitrary representations of GL(D,R): Duality  and
Poincare lemma,''
Commun.\ Math.\ Phys.\  {\bf 245} (2004) 27
[arXiv:hep-th/0208058].
%%CITATION = HEP-TH 0208058;%%

%\cite{Bekaert:2003az}
\bibitem{Bekaert:2003az}
X.~Bekaert and N.~Boulanger,
``On geometric equations and duality for free higher spins,''
Phys.\ Lett.\ B {\bf 561} (2003) 183
[arXiv:hep-th/0301243].
%%CITATION = HEP-TH 0301243;%%
%\cite{Dubois-Violette:1999rd}



\bibitem{Ramaswamy}
S.~Ramaswamy and A.~Sen, ``{Dual-mass in general relativity},'' {\em
  J.Math.Phys.} {\bf 22} (1981)  2612.

\bibitem{Ashtekar}
A.~Ashtekar and A.~Sen, ``{NUT} 4{‐}momenta are forever,'' {\em Journal of
  Mathematical Physics} {\bf 23} (1982) no.~11, 2168--2178.



  \bibitem{Curtright:1980yk}
T.~Curtright, \emph{{Generalized Gauge Fields}},
  %\href{http://dx.doi.org/10.1016/0370-2693(85)91235-3}
  {\emph{Phys. Lett.} {\bf
  B165} (1985) 304}.

\bibitem{Labastida:1986gy}
J.~M.~F. Labastida and T.~R. Morris, \emph{{Massless Mixed Symmetry Bosonic
  Free Fields}},
  %\href{http://dx.doi.org/10.1016/0370-2693(86)90143-7}
  {\emph{Phys. Lett.} {\bf
  B180} (1986) 101}.

\bibitem{Labastida:1986ft}
J.~M.~F. Labastida, \emph{{Massless Bosonic Free Fields}},
%  \href{http://dx.doi.org/10.1103/PhysRevLett.58.531}
{\emph{Phys. Rev. Lett.}
  {\bf 58} (1987) 531}.

\bibitem{Labastida:1987kw}
J.~M.~F. Labastida, \emph{{Massless Particles in Arbitrary Representations of
  the Lorentz Group}},
 % \href{http://dx.doi.org/10.1016/0550-3213(89)90490-2}
 {\emph{Nucl. Phys.} {\bf
  B322} (1989) 185}.



\bibitem{Dubois-Violette:1999rd}
M.~Dubois-Violette and M.~Henneaux,
``Generalized cohomology for irreducible tensor fields of mixed Young  symmetry
type,''
Lett.\ Math.\ Phys.\  {\bf 49} (1999) 245
[arXiv:math.qa/9907135].
%%CITATION = MATH-QA 9907135;%%



 %\cite{Howe:2015bdd}
\bibitem{Howe:2015bdd}
P.~S.~Howe and U.~Lindstr\"om,
``Notes on Super Killing Tensors,''
JHEP \textbf{03} (2016), 078
%doi:10.1007/JHEP03(2016)078
[arXiv:1511.04575 [hep-th]].
%11 citations counted in INSPIRE as of 25 Aug 2022
  %%%%%%%%%%%%%%%%%%%%%%%%%

%\cite{Howe:2018lwu}
\bibitem{Howe:2018lwu}
P.~S.~Howe and U.~Lindstr\"om,
``Some remarks on (super)-conformal Killing-Yano tensors,''
JHEP \textbf{11} (2018), 049
%doi:10.1007/JHEP11(2018)049
[arXiv:1808.00583 [hep-th]].
%14 citations counted in INSPIRE as of 25 Aug 2022



%\cite{Lee:1990nz}
\bibitem{Lee:1990nz}
J.~Lee and R.~M.~Wald,
``Local symmetries and constraints,''
J. Math. Phys. \textbf{31} (1990), 725-743
%doi:10.1063/1.528801
%561 citations counted in INSPIRE as of 23 Aug 2022


%\cite{Wald:1999wa}
\bibitem{Wald:1999wa}
R.~M.~Wald and A.~Zoupas,
``A General definition of 'conserved quantities' in general relativity and other theories of gravity,''
Phys. Rev. D \textbf{61} (2000), 084027
%doi:10.1103/PhysRevD.61.084027
[arXiv:gr-qc/9911095 [gr-qc]].
%465 citations counted in INSPIRE as of 23 Aug 2022




%\cite{Barnich:1994db}
\bibitem{Barnich:1994db}
G.~Barnich, F.~Brandt and M.~Henneaux,
``Local BRST cohomology in the antifield formalism. 1. General theorems,''
Commun. Math. Phys. \textbf{174} (1995), 57-92
%doi:10.1007/BF02099464
[arXiv:hep-th/9405109 [hep-th]].
%355 citations counted in INSPIRE as of 23 Aug 2022

%\cite{Barnich:2001jy}
\bibitem{Barnich:2001jy}
G.~Barnich and F.~Brandt,
``Covariant theory of asymptotic symmetries, conservation laws and central charges,''
Nucl. Phys. B \textbf{633} (2002), 3-82
%doi:10.1016/S0550-3213(02)00251-1
[arXiv:hep-th/0111246 [hep-th]].
%519 citations counted in INSPIRE as of 23 Aug 2022







%\cite{Barnich:2000zw}
\bibitem{Barnich:2000zw}
G.~Barnich, F.~Brandt and M.~Henneaux,
``Local BRST cohomology in gauge theories,''
Phys. Rept. \textbf{338} (2000), 439-569
%doi:10.1016/S0370-1573(00)00049-1
[arXiv:hep-th/0002245 [hep-th]].
%443 citations counted in INSPIRE as of 23 Aug 2022

%\cite{Hull:2023dgp}
\bibitem{Hull:2023dgp}
C.~M.~Hull,
``Covariant Action for Self-Dual p-Form Gauge Fields in General Spacetimes,''
[arXiv:2307.04748 [hep-th]].
%2 citations counted in INSPIRE as of 13 Nov 2023

%\cite{Barnich:2008ts}
\bibitem{Barnich:2008ts}
G.~Barnich and C.~Troessaert,
``Manifest spin 2 duality with electric and magnetic sources,''
JHEP \textbf{01} (2009), 030
%doi:10.1088/1126-6708/2009/01/030
[arXiv:0812.0552 [hep-th]].
%22 citations counted in INSPIRE as of 23 Sep 2022



%\cite{Nester:1981bjx}
\bibitem{Nester:1981bjx}
J.~A.~Nester,
``A New gravitational energy expression with a simple positivity proof,''
Phys. Lett. A \textbf{83} (1981), 241
%doi:10.1016/0375-9601(81)90972-5
%195 citations counted in INSPIRE as of 25 Aug 2022



\bibitem{Hutt}
C.~Hull, M.~Hutt and U.~Lindstrom, in preparation.


 %\cite{Bossard:2008sw}
\bibitem{Bossard:2008sw}
G.~Bossard, H.~Nicolai and K.~S.~Stelle,
``Gravitational multi-NUT solitons, Komar masses and charges,''
Gen. Rel. Grav. \textbf{41} (2009), 1367-1379
%doi:10.1007/s10714-008-0720-7
[arXiv:0809.5218 [hep-th]].
%26 citations counted in INSPIRE as of 25 Aug 2022



%\cite{Deser:1988fc}
\bibitem{Deser:1988fc}
S.~Deser and M.~Soldate,
``Gravitational Energy in Spaces With Compactified Dimensions,''
Nucl. Phys. B \textbf{311} (1989), 739-750
%doi:10.1016/0550-3213(89)90175-2
%35 citations counted in INSPIRE as of 12 Oct 2023



 \bibitem{Newman:1963yy}
E.~Newman, L.~Tamburino and T.~Unti, ``Empty Space Generalization
Of The Schwarzschild Metric,'' J.\ Math.\ Phys.\  {\bf 4}, 915
(1963).
%%CITATION = JMAPA,4,915;%%

\bibitem{Taub}
A.~H. Taub, ``Empty space-times admitting a three parameter group of motions,''
  {\em Annals of Mathematics} {\bf 53} (1951) no.~3, 472--490.


%\cite{Hinterbichler:2022agn}
\bibitem{Hinterbichler:2022agn}
K.~Hinterbichler, D.~M.~Hofman, A.~Joyce and G.~Mathys,
``Gravity as a gapless phase and biform symmetries,''
JHEP \textbf{02} (2023), 151
%doi:10.1007/JHEP02(2023)151
[arXiv:2205.12272 [hep-th]].
%10 citations counted in INSPIRE as of 24 Mar 2023


%\cite{Hull:1983ap}
\bibitem{Hull:1983ap}
C.~M.~Hull,
``The Positivity of Gravitational Energy and Global Supersymmetry,''
Commun. Math. Phys. \textbf{90} (1983), 545
%doi:10.1007/BF01216185
%74 citations counted in INSPIRE as of 25 Aug 2022


%\cite{Bombelli:1986sb}
\bibitem{Bombelli:1986sb}
L.~Bombelli, R.~K.~Koul, G.~Kunstatter, J.~Lee and R.~D.~Sorkin,
``On Energy in Five-dimensional Gravity and the Mass of the {Kaluza-Klein} Monopole,''
Nucl. Phys. B \textbf{289} (1987), 735-756
%doi:10.1016/0550-3213(87)90404-4
%40 citations counted in INSPIRE as of 25 Oct 2023



%\cite{Nguyen:2022nnx}
\bibitem{Nguyen:2022nnx}
K.~Nguyen and P.~West,
``Conserved asymptotic charges for any massless particle,''
[arXiv:2208.08234 [hep-th]].
%0 citations counted in INSPIRE as of 25 Aug 2022
%%%%%


%\cite{Boulanger:2015mka}
\bibitem{Boulanger:2015mka}
N.~Boulanger, P.~Sundell and P.~West,
``Gauge fields and infinite chains of dualities,''
JHEP \textbf{09} (2015), 192
%doi:10.1007/JHEP09(2015)192
[arXiv:1502.07909 [hep-th]].
%29 citations counted in INSPIRE as of 25 Oct 2023




%\cite{Penrose:1982wp}
\bibitem{Penrose:1982wp}
R.~Penrose,
``Quasilocal mass and angular momentum in general relativity,''
Proc. Roy. Soc. Lond. A \textbf{381} (1982), 53-63
%doi:10.1098/rspa.1982.0058
%276 citations counted in INSPIRE as of 25 Oct 2023

%\cite{Benedetti:2021lxj}
\bibitem{Benedetti:2021lxj}
V.~Benedetti, H.~Casini and J.~M.~Magan,
``Generalized symmetries of the graviton,''
JHEP \textbf{05} (2022), 045
%doi:10.1007/JHEP05(2022)045
[arXiv:2111.12089 [hep-th]].
%5 citations counted in INSPIRE as of 23 Sep 2022

%\cite{Benedetti:2023ipt}
\bibitem{Benedetti:2023ipt}
V.~Benedetti, P.~Bueno and J.~M.~Magan,
``Generalized Symmetries for Generalized Gravitons,''
Phys. Rev. Lett. \textbf{131} (2023) no.11, 111603
%doi:10.1103/PhysRevLett.131.111603
[arXiv:2305.13361 [hep-th]].
%3 citations counted in INSPIRE as of 25 Oct 2023

%\cite{Gomez-Fayren:2023qly}
\bibitem{Gomez-Fayren:2023qly}
C.~G\'omez-Fayr\'en, P.~Meessen and T.~Ort\'\i{}n,
``Covariant generalized conserved charges of General Relativity,''
JHEP \textbf{09} (2023), 174
%doi:10.1007/JHEP09(2023)174
[arXiv:2307.04041 [gr-qc]].
%0 citations counted in INSPIRE as of 25 Oct 2023


%\cite{Godazgar:2018qpq}
\bibitem{Godazgar:2018qpq}
H.~Godazgar, M.~Godazgar and C.~N.~Pope,
``New dual gravitational charges,''
Phys. Rev. D \textbf{99} (2019) no.2, 024013
%doi:10.1103/PhysRevD.99.024013
[arXiv:1812.01641 [hep-th]].
%53 citations counted in INSPIRE as of 23 Aug 2022



\end{thebibliography}
\end{document}